%% file: main.tex
\documentclass[aps,pra,reprint,superscriptaddress]{revtex4-1}
\pdfoutput=1
\usepackage{graphicx}

\usepackage{amsmath}
\usepackage{amssymb}
\usepackage{bm}
\usepackage{bbm}
\usepackage{braket}
\usepackage{textcomp}
\usepackage[utf8]{inputenc}
\usepackage{multirow}

\newcommand{\vect}[1]{\mathbf{#1}}

\begin{document}

\title{Coherent rotations of qubits within a multi-species ion-trap quantum computer}

\author{Martin W. van Mourik}
\author{Esteban A. Martinez}
\affiliation{Institut f\"ur Experimentalphysik, Universit\"at Innsbruck, Technikerstraße 25/4, 6020 Innsbruck, Austria}
\author{Lukas Gerster}
\author{Pavel Hrmo}
\author{Thomas Monz}
\author{Philipp Schindler}
\affiliation{Institut f\"ur Experimentalphysik, Universit\"at Innsbruck, Technikerstraße 25/4, 6020 Innsbruck, Austria}
\author{Rainer Blatt}
\affiliation{Institut f\"ur Experimentalphysik, Universit\"at Innsbruck, Technikerstraße 25/4, 6020 Innsbruck, Austria}
\affiliation{Institut f\"ur Quantenoptik und Quanteninformation, \"Osterreichische Akademie der Wissenschaften, Technikerstraße 21a, 6020 Innsbruck, Austria}

\begin{abstract}  
\input{abstract}
\end{abstract}

\maketitle

\tableofcontents

\section{Introduction}
\label{sec:introduction}
\input{introduction}

\section{Experimental overview}
\label{sec:exp_overview}
\input{exp_overview}

\section{Rotation potentials}
\label{sec:trap_potentials}
\input{potentials}

\section{Experimental implementation}
\label{sec:ion_rotations}
\input{ion_rotations}

\section{Experimental results}
\label{sec:exp_results}
\input{results}

\section{Outlook}
\label{sec:outlook}
\input{outlook}

\section*{Acknowledgements}
\label{sec:acknowledgements}
\input{acknowledgements}

\appendix
\section{Equilibrium position for two ions}
\label{sec:appendix_equil_pos}
\input{appendix_equil_pos}

\section{Voltage solutions for spherical harmonic potentials}
\label{sec:appendix_voltagesolutions}
\input{appendix_voltage_solutions.tex}

\section{2-Ion heating measurement}
\label{sec:appendix_JC}
\input{appendix_JC}

\input{main.bbl}
\end{document}

%% file: abstract.tex
We describe, realize, and experimentally investigate a method to perform physical rotations of ion chains, trapped in a segmented surface Paul trap, as a building block for large scale quantum computational sequences. Control of trapping potentials is achieved by parametrizing electrode voltages in terms of spherical harmonic potentials. Voltage sequences that enable crystal rotations are numerically obtained by optimizing time-dependent ion positions and motional frequencies, taking into account the effect of electrical filters in our set-up. We minimize rotation-induced heating by expanding the sequences into Fourier components, and optimizing the resulting parameters with a machine-learning approach. Optimized sequences rotate $^{40}$Ca$^+$ - $^{40}$Ca$^+$ crystals with axial heating rates of $\Delta\bar{n}_{com}=0.6^{(+3)}_{(-2)}$ and $\Delta\bar{n}_{str}=3.9(5)$ phonons per rotation for the common and stretch modes, at mode frequencies of 1.24 and 2.15 MHz. Qubit coherence loss is 0.2(2)$\%$ per rotation. We also investigate rotations of mixed species crystals ($^{40}$Ca$^+$ - $^{88}$Sr$^+$) and achieve unity success rate.

%% file: introduction.tex
Trapped-ion based systems are a promising platform for quantum computation \cite{Ladd_Nature_2010}, as they benefit from long coherence times \cite{Wang_NaturePhotonics_2017}, high entangling gate fidelities \cite{Ballance_PhysRevLett_2016,Gaebler_PhysRevLett_2016}, and quick and high-fidelity readout \cite{Myerson_PhysRevLett_2008}. While these key ingredients have been demonstrated in macroscopic Paul traps, the 3-dimensional architecture makes scalability technically challenging. For example, as the number of trapped ions $N$ in a crystal increases, maintaining a linear crystal chain of individually addressable ions requires low trapping frequencies, where qubits are more sensitive to external noise \cite{Turchette_PhysRevA_2000}, while gate times increase as $\sqrt{N}$. Furthermore, a higher number of ions introduces linearly more motional modes, such that frequency crowding between motional modes becomes inevitable \cite{Brown_Nature_2016}. The quality of gate operations can be maintained by employing multiple trapping regions, thus avoiding long ion strings and frequency crowding per computational region \cite{Kielpinski_Nature_2002}. Physically separating parts of a quantum register requires the trap electrodes to be divided into segments \cite{Hughes_PhysRevLett_1996}. A Paul trap can incorporate segmented trap electrodes and can be designed to include junctions \cite{Hensinger_APL_2006}, allowing for a wide range of possible trap designs. While it is possible to fabricate 3D segmented traps, it is technically challenging \cite{Hughes_ContPhys_2011}. The fabrication of planar (2D) traps benefits from an already well-established field of surface lithography \cite{Mack_2008}, making them versatile for trap design when considering architectures for large scale quantum computation \cite{Seidelin_PhysRevLett_2006, Mehta_ApplPhysLett_2014}.

For ions in a single crystal, interactions between any subset of ions can be achieved by single ion addressing, for example by spectroscopically decoupling non-interacting ions \cite{Schindler_NJP_2013}.
As the number of ions in a crystal increases, so does the amount of required decoupling pulses. Thus, the induced error of this method poses an additional restriction when considering scalability.
Segmented electrodes have the advantage that the spectroscopic operations used for decoupling ions can be replaced with physical manipulation of an ion string, such as shuttling \cite{Walther_PhysRevLett_2012}, splitting \cite{Kaufmann_NJP_2014}, and merging.
Such operations can be achieved in time scales on the order of the ions' motional frequencies, typically several microseconds \cite{Bowler_PhysRevLett_2012}. No lasers are required, and the qubit states are left unaffected.

Shuttling and splitting operations alone, however, are not sufficient for interactions beyond those of nearest-neighboring ions. It was proposed that one could achieve long-distance interactions by reordering ion crystals through the use of junctions \cite{Kielpinski_Nature_2002}. Due to the complicated electrode structure required to generate the radio-frequency field to confine the ions in the junction region, shuttling individual ions through a junction is challenging and typically results in unwanted disturbance (heating) of the crystal's motional state \cite{Shu_PhysRevA_2014,Blakestad_PhysRevLett_2009}, at hundreds of microseconds transfer durations \cite{Blakestad_PhysRevA_2011}. Since optical state manipulation is dependent on the motional state, these disturbances are detrimental to the fidelity of subsequent gates and should be avoided. Another approach to reordering the ion crystal in a linear trapping architecture is by rotating ions around each other within a single trapping region. This has the advantages of not passing through a junction, with the potential of showing less motional mode heating compared to heating in junctions, and to simplify trap design.

The feasibility of ion reconfiguration has been demonstrated in segmented traps through junctions \cite{Blakestad_PhysRevLett_2009,Wright_NJP_2013} and in 3D traps using rotations \cite{Kaufmann_PRL_2017}.
Rotations on a planar trap, however, come with more challenges compared to 3D traps \cite{Splatt_NJP_2009,Maunz_2017}: since the trapping region is typically above the surface of the trap, the trap electrodes cannot be placed symmetrically around the trapping region. Therefore, supplying voltages that keep ions at the trap center requires an accurate model of the potentials generated by trap electrodes. Furthermore, surface traps have a higher degree of anharmonicity, and typically an order of magnitude lower trap depth \cite{Chiaverini_QIC_2005,Seidelin_PhysRevLett_2006}. For these reasons, the requirement to keep ions near the trap center of a surface trap is more stringent than in 3D traps.

In this work we investigate ion crystal rotations in a planar trapping architecture. We discuss the framework required for determining electrode voltage sequences that facilitate rotations. Due to the higher level of complexity in controlling potentials in a surface trap, we focus on methods of simulating, calibrating, and optimizing potentials. Experimentally, we aim to minimize rotation-induced motional heating and maintain coherence, with the prospect of using rotations in quantum computational algorithms. Furthermore, we investigate rotations of multi-species ion crystals, as many proposed algorithms in quantum computation \cite{Bermudez_PhysRevX_2017} benefit from the use of multiple ion species.

This work is structured as follows: Section \ref{sec:exp_overview} gives an overview of the employed experimental setup and a qualitative description of ion crystal rotations in a planar traps. Section \ref{sec:trap_potentials} provides the groundwork for calculating the potentials required to rotate ion crystals, and determining trap electrode voltages to realize these potentials. Section \ref{sec:ion_rotations} describes the application of the methods described in Section \ref{sec:trap_potentials}. Section \ref{sec:exp_results} describes success rate, mode heating, and qubit coherence after rotation sequences. Section \ref{sec:outlook} concludes with a summary and outlook.

%% file: exp_overview.tex
\begin{figure}
	\includegraphics[width=\linewidth]{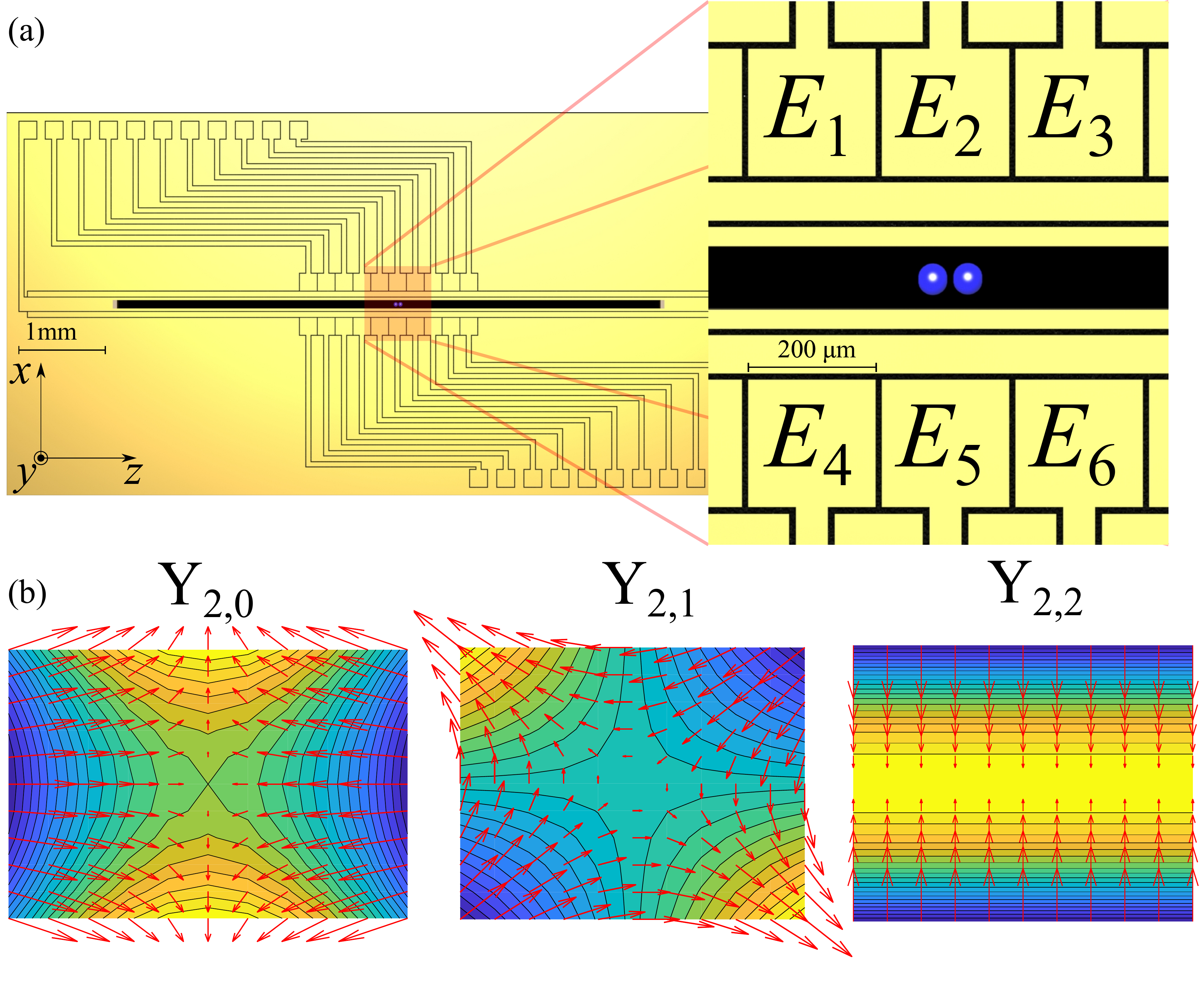}
	\caption{(a) Schematic of the planar segmented ion trap. (b) Equipotential contours of spherical harmonic potential terms $Y_{2,0}$, $Y_{2,1}$, and $Y_{2,2}$, parametrically used to generate a controlled rotation. Note that in addition to these applied potentials, the RF pseudo-potential is also present.}
	\label{fig:Multipole_Potentials}
\end{figure}

The experiments have been carried out with $^{40}$Ca$^{+}$ and $^{88}$Sr$^{+}$ ions, on a planar slotted segmented trap, in a cryogenic environment \cite{Brandl_RSI_2016}. The trap, depicted in Figure~\ref{fig:Multipole_Potentials}, consists of 10 individually addressable electrode pairs, a central electrode and central RF rails. The electrode pairs are 200 $\mu$m wide, and are separated by 10 $\mu$m wide trenches. Voltages applied to each electrode are filtered by a second-order LC filter, with a cutoff frequency of 47 kHz. Ions are trapped 110 $\mu$m above the trap plane above a 100 $\mu$m wide slot, yielding a 121 $\mu$m ion-electrode separation. Ion crystals are oriented along the $z$-axis, and the RF rails produce a confining pseudopotential in the $xy$-plane. The trap is depicted in Figure \ref{fig:Multipole_Potentials}(a), with the electrodes closest to the trapping region labeled $E_1 - E_6$.

An intuitive illustration of the rotation process is as follows: a sequence of electrostatic potentials is applied to the trapping region, by means of voltages on the trap's DC electrodes. A positive(negative) bias voltage is applied to opposite corner electrodes, $E_3,E_4$($E_1,E_6$), which tilts the ion chain from it's original trap axis. The confinement of the weakest trapping axis ($z$) is increased by increasing the voltage on corner electrodes, and decreasing the voltage on the central electrodes $E_2,E_5$, until the ions are aligned along the $x$-axis. The polarity of the diagonal bias voltages of the first step is then reversed, and the along-axis confining potential is relaxed. After removing the diagonal bias, the ions are back in their original equilibrium positions, and have undergone a rotation.

Naively applying voltages on electrodes as prescribed above typically results in additional undesired electric fields, thus displacing the ions from their equilibrium position, inducing excessive micromotion, potentially enough to expel ions from the trap. Improved control is therefore necessary, and is achieved by parametrising the potential generated around the trapping region by each electrode, thus giving control in terms of trapping potentials instead of in terms of electrode voltages. We choose spherical harmonics \cite{Jackson_1999} as parameters, since they contain potentials that are natural for ion trapping and manipulation, and can be controlled independently. These potentials, denoted by $Y_{l,n}(x,y,z)$, with the degree $l$ and order $n$, are a function of position and can be analytically expressed in Cartesian coordinates.

Any static field potential $\phi$ can be approximated up to a chosen degree as a linear combination of these spherical harmonics, as $\phi\approx\sum_{l,n}m_{l,n}Y_{l,n}$, with $m_{l,n}$ being scalar coefficients.
The five second-order terms together fully define any 3-dimensional quadratic potential (with the restriction that $\nabla^2\phi=0$). Expanding and controlling the potential up to second-order spherical harmonics is therefore sufficient to describe the potentials that are required for ion trapping, shuttling, and rotations. 
The (normalized) spherical harmonic potentials are given, up to second degree, by:
\begin{equation}
\label{eq:spherical_harmonics}
\begin{aligned}
\lbrace Y_{1,-1}, Y_{1,0}, Y_{1,1}\rbrace &=\sqrt{\frac{3}{4\pi}}\lbrace y,z,x\rbrace \\
Y_{2,-2} &= \frac{1}{2}\sqrt{\frac{15}{\pi}} xy\\
Y_{2,-1} &= \frac{1}{2}\sqrt{\frac{15}{\pi}}yz\\
Y_{2,0} &= \frac{1}{4}\sqrt{\frac{5}{\pi}}(2z^2 - x^2 - y^2)\\
Y_{2,1} &= \frac{1}{2}\sqrt{\frac{15}{\pi}}xz\\
Y_{2,2} &= \frac{1}{4}\sqrt{\frac{15}{\pi}}(x^2 - y^2)
\end{aligned}
\end{equation}
We choose to define $z$ to be the direction along the trap axis, and $x$ and $y$ to be in the parallel and perpendicular radial directions, with respect to the trap plane. This expansion in spherical harmonics and choice of axes is convenient for ion trapping: spherical harmonics include a term that produces a potential that confines ions along the axial direction ($Y_{2,0}$), similar to end-caps in 3D Paul traps, terms that produce homogeneous fields ($Y_{1,-1}$, $Y_{1,0}$, $Y_{1,1}$), and several terms for producing diagonal offsets required for applying rotations. The potential terms $Y_{2,0}$, $Y_{2,1}$, and $Y_{2,2}$, displayed in Figure \ref{fig:Multipole_Potentials}(b), are the main contributors to controlled rotations in the $xz$-plane. It should be noted that for splitting an ion crystal, higher order terms are necessary \cite{Kaufmann_NJP_2014,Home_QuantumInfoComp_2004,Nizamani_ApplPhysB_2012,Eble_OptSocofAmerica_2010}.

We use a numerical electrostatic calculation of the trap potentials to determine the voltages that are required to control the individual spherical harmonic potentials. Tolerances in manufacturing the trap limit the accuracy of these simulations and thus we calibrate potentials as discussed in Section \ref{sec:potential_calibration}.
Voltages are applied to individual electrodes by a custom arbitrary waveform generator (AWG), similar to that used in Ref \cite{Ruster_PhysRevA_2014}, supplied by the team of Prof. F. Schmidt-Kaler, at the Johannes Gutenberg University Mainz.

%% file: potentials.tex
A given set of electrode voltages generates a static potential, which, together with the RF pseudo-potential, determines an ion chain's position, spacing, and mode frequencies. To perform rotations, one needs to apply a sequence of voltages to the trap's electrodes to control the full trap potential. In this section we describe a model for the trapping potentials and corresponding trap frequencies, and then the sequence of potentials required for generating ion crystal rotations.

\subsection{Rotation potentials}
\label{sec:Rotation_potentials}
The total potential energy $E_{pot}$ of a trapped ion crystal is given by the trap's potential energy $E_{trap}$ and the interaction energy of the ions $E_{coul}$,
\begin{equation}
E_{pot}=E_{trap}+E_{coul}.
\end{equation}
The Coulomb energy for $N$ ions of charge $q_i$ and at positions $\textbf{x}^{(i)}$ is:
\begin{equation}
E_{coul}=\frac{1}{4\pi\epsilon_0}\sum_i^N\sum_{j<i}^N \frac{q_i q_j}{\left|\textbf{x}^{(j)}-\textbf{x}^{(i)}\right|}.
\end{equation}
The potential energy of the trap is approximated by a harmonic RF component $\Phi_{RF}$ in the pseudo-potential approximation (thus unique for ions of different masses and charges), a harmonic DC component $\Phi_{DC}$, and a linear DC component $\Phi_{E}$. The RF and DC harmonic potentials can each be described by a $3\times 3$ curvature matrix ($V_{RF}^{(i)}$ and $V_{DC}$, respectively), which are the Hessian of $\Phi_{RF}$ and $\Phi_{DC}$, and contain only scalar terms. The potential $\Phi_{E}$, originating from an undesired homogeneous electric field, can be expressed by its electric field vector, $\textbf{E}=-\nabla\Phi_{E}$.

The trap potential is given by:
\begin{widetext}
\begin{equation}
\label{eq:trap_potential}
E_{trap}=\frac{1}{2}\sum_i^N q_i \textbf{x}^{(i)T} V_{RF}^{(i)}\textbf{x}^{(i)}+\frac{1}{2}\sum_i^N q_i\textbf{x}^{(i)T}V_{DC}\textbf{x}^{(i)} 
-\sum_i^N q_i\textbf{E}\cdot\textbf{x}^{(i)},
\end{equation}
\end{widetext}
This model assumes that a local quadrupole field is generated around $\textbf{x}=\textbf{0}$ by both $V_{RF}$ and $V_{DC}$.
In order to satisfy Maxwell's equation $\nabla^2 \Phi_{DC}=0$ for static fields, it is required that $V_{DC}$ satisfies $tr\lbrace V_{DC}\rbrace=0$. The RF pseudo-potential does not obey $\nabla^2 \Phi_{RF}=0$. Assuming that the trap's RF potential does not have a curvature in one dimension (the trap's axial direction), the coordinate basis can be chosen such that the pseudo-potential's curvature matrix at its minimum point is given by:
\begin{equation}
    V_{RF}^{(i)}=\frac{(q^{(i)})^2}{4m^{(i)}\Omega^2_{RF}}\left|\nabla\Phi_{RF}\right|^2_{min}\begin{pmatrix}
    1 & 0 & 0 \\
	0 & 1 & 0 \\
    0 & 0 & 0
    \end{pmatrix}
\end{equation}
with $m$ an ion's mass, and $\Omega_{RF}$ the RF drive frequency. We denote the factor in front of the matrix as $v^{(i)}_{RF}$.

Any static quadrupole term $V_{DC}$ can be defined by a linear combination of the orthogonal basis,
\begin{align}
    V_{DC}^{(l)}=
    \left\lbrace\begin{pmatrix} 
	0 & 1/2 & 0 \\
	1/2 & 0 & 0 \\
    0 & 0 & 0 
	\end{pmatrix}
    ,
    \begin{pmatrix}
    0 & 0 & 0 \\
	0 & 0 & 1/2 \\
    0 & 1/2 & 0 
    \end{pmatrix}\right.
    \\
    \left.\begin{pmatrix}
    -1 & 0 & 0 \\
	0 & -1 & 0 \\
    0 & 0 & 2 
    \end{pmatrix}
    ,
    \begin{pmatrix}
    0 & 0 & 1/2 \\
	0 & 0 & 0 \\
    1/2 & 0 & 0 
    \end{pmatrix}
    ,
    \begin{pmatrix}
    1 & 0 & 0 \\
	0 & -1 & 0 \\
    0 & 0 & 0 
    \end{pmatrix}\right\rbrace \nonumber
\end{align}
as $V_{DC}=\sum_l v_l V_{DC}^{(l)}$. Here, $v_l$ are coefficients, grouped in the vector $\textbf{v}$.
This basis encompasses the second degree spherical harmonic potentials in Cartesian coordinates $Y_{2,\lbrace-2,-1,0,1,2\rbrace}$. 

Grouping all individual ion positions $\textbf{x}^{(i)}$ into a single vector $\textbf{x}$ of length $3N$, the equilibrium positions $\textbf{x}_0$ of all ions can be determined by solving the set of differential equations
\begin{equation}
\label{eq:equilibrium_positions}
\frac{\partial E_{pot}}{\partial\textbf{x}}=\mathbf{0}.
\end{equation}

Motional modes and their energies (and thus frequencies $\omega_{\lbrace x,y,z\rbrace}^{(i)}$) can be obtained by finding the eigenvectors and eigenvalues of the mass-corrected Hessian matrix $H$, with masses $m_i$, given by the terms
\begin{equation}
\label{eq:frequencies}
H_{ij}=\frac{1}{\sqrt{m_i m_j}}\left. \frac{\partial^2 E_{pot}}{\partial x_i\partial x_j}\right|_{\textbf{x}_0},
\end{equation}
linearized around $\textbf{x}_0$. Real positive eigenvalues indicate a stable trapping potential.

Using this model, we find solutions of the vector $\textbf{v}^{(n)}$, at discrete steps $n$, that facilitate the rotation of an ion crystal. We keep the value of $v^{(i)}_{RF}$ fixed, though this could potentially be included as a tunable parameter. A suitable starting potential $\textbf{v}^{(0)}$ is characterized by stable trapping frequencies, in our trap on the order of $\omega_{\left\lbrace x,y,z\right\rbrace}\approx2\pi\left\lbrace3,3,1\right\rbrace$ MHz, though avoiding degeneracy in $\omega_x$ and $\omega_y$ by about 200 kHz. This corresponds to $v_3^{(0)} \approx 10^7$ V/m$^2$ and $v_{RF}^{(i)} \approx 1.6\times10^8$ V/m$^2$. Setting $v_5^{(0)} \approx10^7$ V/m$^2$ ensures that degeneracy in $\omega_x$ and $\omega_y$ is lifted.

The goal is to numerically find values for $\textbf{v}^{(n)}$ that minimize a cost function $C(\textbf{v})$ at steps $n$ of an ion crystal rotation, characterized by a rotation angle $\theta_n$. Here, $\theta_n\in [0,\pi]$ is the angle between two ions in a chain and the $z$-axis, in the $xz$-plane, at step $n$. The cost function is foremost composed of the restriction that $\textbf{v}$ should place the ions at the required angle $\theta_n$. Additional restrictions are imposed to improve the quality of the operations, noting that we aim to minimize mode heating. In this scope, high mode frequencies during a rotation are favorable \cite{Bruzewicz_PhysRevA_2015}. Furthermore, degeneracy in mode frequencies induces crosstalk in the modes of motion making undesired exchange of phonons across different modes more likely. After a rotation, the potential should be identical to the initial potential.

Placing ions at the required angle $\theta_n$ is achieved by minimizing the cost function
\begin{equation}
C_0 (\textbf{v}^{(n)}) = 1-
\frac{(\sin\theta_n,0,\cos\theta_n)\cdot(\textbf{x}^{(j)}_0-\textbf{x}^{(i)}_0)}
{\left|\textbf{x}^{(j)}_0-\textbf{x}^{(i)}_0\right|},
\end{equation}
which parametrizes the potential required for a two-ion ($i$ and $j$) crystal rotation.
We use the solution of $\textbf{v}$ at $\theta_{n-1}$ as an initial value for the minimization of $C_0 (\textbf{v})$ at $\theta_{n}$, to reduce computation time and ensure a smooth rotation. For a single-species two-ion crystal, the equilibrium positions $\textbf{x}^{(0)}$ can be analytically calculated (see Appendix \ref{sec:appendix_equil_pos}), thus reducing computational cost.

Degeneracy in mode frequencies $\omega_i$ is avoided with the cost function
\begin{equation}
C_1 (\textbf{v})=\sum_{i} \sum_{j>i} \frac{1}{(\omega_i-\omega_j)^2}.
\end{equation}
Maintaining high frequencies can be implemented by the cost function
\begin{equation}
C_2 (\textbf{v})=\sum_i \frac{1}{\omega_i^2}
\end{equation}
Ensuring that the final potential (at $\theta=180^{\circ}$) is identical to the initial potential ($\theta=0$), can be attained by keeping frequencies as close as possible to their initial values $\omega_i^{(0)}$ throughout the entire rotation, using the cost function
\begin{equation}
C_3 (\textbf{v})=\sum_i (\omega_i-\omega_i^{(0)})^2
\end{equation}
An indirect consequence of $C_3$ is that it helps achieve the conditions imposed by $C_1$ and $C_2$.
The full cost function to be minimized is given by
\begin{equation}
C(\textbf{v})=\sum_{i=0}^3 \lambda_i C_i(\textbf{v})
\end{equation}
with $\lambda_i$ a set of weights.

For single species rotations, there exists a unique solution that keeps all mode frequencies constant throughout the rotation sequence, for given starting parameters. Setting $\lambda_0$ and $\lambda_3$ to a non-zero value is sufficient for determining this solution uniquely. Multi-species rotations cannot maintain constant frequencies and might thus benefit from $\lambda_1$ and $\lambda_2$ contributions. In the multi-species case, we determine the values of the weights $\lambda_i$ empirically. We find that setting $\lambda_0=10\lambda_1=10\lambda_2=\lambda_3$ produces potential sequences with no discontinuities that uphold the imposed restrictions.

The calculated potentials will be generated by applying voltages to our trap's electrodes. We use boundary-element methods \cite{Singer_RevModPhys_2009} to determine the potentials generated by applying a voltage to each electrode separately. From this, we can determine which set of voltages produce each potential $Y_{l,n}$. Our method for determining electrode voltages to realize trapping potentials is discussed in more detail in appendix \ref{sec:appendix_voltagesolutions}.

\subsection{Potential calibration}
\label{sec:potential_calibration}
The solutions for the required electrode voltages for a given static trapping potential, as derived in appendix \ref{sec:appendix_voltagesolutions}, may not produce the actual expected potentials, due to numerical simulation errors, uncontrolled electric charges on the trap surface, and manufacturing tolerances. These discrepancies need to be estimated to produce optimal rotation potentials. Here, we will introduce a protocol to calibrate the variation between the actual potential and the desired potential.

For a desired potential, expressed in spherical harmonic terms $\textbf{m}_{set}$, we calculate the expected set of electrode voltages $\bm{\alpha}$, following the method described in appendix \ref{sec:appendix_voltagesolutions}. In the experiment, the set of applied voltages, result in a potential $\textbf{m}$ that might deviate from $\textbf{m}_{set}$, which we model as
\begin{equation}
\label{eq:calibration_model}
\textbf{m}=\textbf{A}\textbf{m}_{set} + \textbf{b}
\end{equation}
where $\textbf{A}$ is an $8\times8$ matrix that produces a linear correction, and the vector $\textbf{b}$ corresponds to a constant offset. A method to estimate $\textbf{A}$ and $\textbf{b}$ is outlined below. 

The applied RF power and the eight terms of the vector $\textbf{m}$ uniquely define a potential with ellipsoidal equipotential surfaces. Such an ellipsoid can be characterized by nine parameters: the displacement of the center of the ellipsoid in three Cartesian directions, three tilt angles of the ellipsoid's principle axes with respect to the Cartesian axes, and three lengths of the principle axes of the ellipsoid. These parameters translate into measurable quantities of a single trapped ion, outlined below. The calibration terms $\textbf{A}$ and $\textbf{b}$ can therefore be found by experimentally estimating the actual trapping potential for various multipole settings $\textbf{m}_{set}$, and comparing experimental results to calculated expected potentials. We will now describe how each of the potential parameters are experimentally determined. 

The lengths of the principal axes of an equipotential ellipsoid define the curvature of the potential in three orthogonal directions $V_{\{1,2,3\}}$, and are thus determined by the motional frequencies of the ion, as $\omega_{i}=\sqrt{2V_i q/m}$, with $q$ and $m$ the charge and mass of the ion. We use sideband spectroscopy \cite{Leibfried_RevModPhys_2003} to determine the ion's motional frequencies $\omega_{i,j}$ for various multipole settings $\textbf{m}_{set,j}$, indexed by $j$.

The tilt of the principal axes can be determined by measuring the relative coupling strengths of a beam with wavevector $\textbf{k}$ on resonance with the motional sidebands of an ion's electronic transition. The coupling strength $\Omega_i$ of mode $i$ is dependent on the angle of incidence of the beam with respect to orientation of the motional mode, $\Omega_i\propto \textbf{k}\cdot \hat{\textbf{i}}'$. Here $\hat{\textbf{i}}'=\lbrace \hat{\textbf{x}}',\hat{\textbf{y}}',\hat{\textbf{z}}'\rbrace$ are the unit vectors of the potential's principal axes, accented to denote that they are not necessarily the Cartesian axes. The coupling strength also depends on the motional mode frequency, $\Omega_i\propto \omega^{-1/2}$. One can find the coupling strengths by measuring the effective Rabi frequency when exciting the blue sideband for each motional mode of an ion prepared in the motional and electronic ground state. The tilt of the principal axes can be found by determining their unit vectors by solving the following set of equations:
\begin{eqnarray}
\hat{\textbf{x}}'\times\hat{\textbf{y}}'=&\hat{\textbf{z}}'\\
\frac{\Omega_x}{\Omega_y}=&\frac{\hat{\textbf{k}}\cdot\hat{\textbf{x}}'}{\hat{\textbf{k}}\cdot\hat{\textbf{y}}'}\sqrt{\frac{\omega_y}{\omega_x}}\\
\frac{\Omega_y}{\Omega_z}=&\frac{\hat{\textbf{k}}\cdot\hat{\textbf{y}}'}{\hat{\textbf{k}}\cdot\hat{\textbf{z}}'}\sqrt{\frac{\omega_z}{\omega_y}}
\end{eqnarray}
Using this method assumes that the wavevector $\textbf{k}$ is known. If there is uncertainty in the wavevector, one can also compare the sideband coupling strengths to that of a carrier excitation $\Omega_c$, which gives the following additional equations:
\begin{equation}
\frac{\Omega_i}{\Omega_c} = \sqrt{\frac{\hbar}{2m\omega_i}}\hat{\textbf{k}}\cdot\hat{\textbf{i}}'
\end{equation}

Finally, a displacement of the ellipsoid, caused by a uniform field, results in an identical displacement of an ion, $\textbf{u}$. Therefore, one can measure the position of the ion to characterize this component of the potential. In our experiment, a CCD camera monitors the position of a single ion with respect to the trap plane. We thus determine the displacement in the xz-plane, denoted by $u_x$ and $u_z$, for various multipole settings $\textbf{m}$ by monitoring the position of an ion on the CCD readout. The CCD image magnification is calibrated by using imaged trap electrodes as a scale reference. Alternatively, one can calculate the separation $d$ of two trapped ions for a measured common mode frequency $\omega$ with $d^3 = q^2/(2\pi\epsilon_0 m\omega^2$), and compare this to the separation on the CCD ($q$ and $m$ are a single ion's charge and mass, and $\epsilon_0$ the vacuum permittivity).

The displacement $u_y$ is perpendicular to the image plane, and thus cannot be detected by the CCD. A trapped ion exhibits motion driven by the time-dependent RF potential, termed micromotion. The amplitude of this motion depends linearly on the displacement of the ion from the RF pseudopotential minimum. We can therefore infer ion displacement by measuring the micromotion amplitude. This amplitude is measured with micromotion sideband spectroscopy \cite{Keller_JAP_2015}, in which we determine the ratio between the coupling strength of a micromotion sideband $\Omega_{MM}$, and that of a carrier transition $\Omega_{car}$. The micromotion amplitude, denoted by the micromotion modulation index $\beta$, is given by $\beta/2\approx\Omega_{MM}/\Omega_{car}$. The modulation index $\beta$ is related to ion displacement as $u_y=2\beta/kq_y$ \cite{Berkeland_JApplPhys_1998}. Here, $k$ is the wavenumber of a beam propagating perpendicular to the trap surface, and $q_y$ is the trap's stability parameter \cite{Leibfried_RevModPhys_2003}.

For any given multipole setting $\textbf{m}_j$, positions $\tilde{u}_i(\textbf{m}_j)$ can be calculated using equation \ref{eq:equilibrium_positions}, and frequencies $\tilde{\omega}_i(\textbf{m}_j)$ and tilts $\tilde{\hat{\textbf{i}}}(\textbf{m}_j)$ are given by the eigenvalues and eigenvectors of equation \ref{eq:frequencies}.
Calculated values of these parameters for various intended multipoles $\textbf{m}_j$ are compared to measured values, giving a cost function given by:
\begin{align}
\label{eq:calibration_minimization}
& \sum_{i,j}(\omega_{i,j}-\tilde{\omega}_i(\textbf{m}_j))^2\nonumber \\
+& \sum_{i,j}\left|\hat{\textbf{i}}_j' - \tilde{\hat{\textbf{i}}}(\textbf{m}_j)\right|^2 \nonumber \\
+& \sum_{i,j}(u_{i,j}-\tilde{u}_i(\textbf{m}_j))^2
\end{align}
where the sum is taken over all measurements $j$. In our model the actual multipole potential $\textbf{m}_j$ depends on the applied multipole setting $\textbf{m}_{set,j}$, given by equation \ref{eq:calibration_model}. We thus find optimized values for $\textbf{A}$ and $\textbf{b}$ by minimizing eq. \ref{eq:calibration_minimization}.

Despite meticulous calibration for static potentials, in our experiment potentials deviate from set potentials during dynamic rotation sequences (further discussed in sections on electrical filters, Section \ref{subsec:filters}).
Noting this deviation, we elect to simplify our calibration such that only the offset field and the motional frequencies are calibrated. This simplification makes the optimization of $\textbf{A}$ underdetermined. We therefore set the off-diagonal elements of $\textbf{A}$ that contribute to the tilt of the potential to 0 (This corresponds to the fourth, fifth, and seventh rows of $\textbf{A}$).
Since this restriction on $\textbf{A}$ only affects the calibration of the tilt of the potential, this method remains an accurate predictor of motional frequencies for any set potential $\textbf{m}_{set}$.

We are able to calibrate our model such that individual errors in frequencies $\left|1-(\omega_{i,j}-\tilde{\omega}_{i}(\textbf{m}_j))/\omega_{i,j}\right|$ are less than 1\%. We have performed further sideband spectroscopy for 2-ion crystals, single and mixed species, for various multipole settings, and verified that the calibrated model is able to predict all trap frequencies within 1\% for a single species crystal, and within 2\% for multi-species crystals.

%% file: ion_rotations.tex
\subsection{General principle}
\label{sec:rotation_principle}
The procedure for crystal rotations is conceptually similar to the rotations applied in a 3-dimensional segmented trap by Kaufmann et al. \cite{Kaufmann_PRL_2017}. In 3D traps, electrodes are typically placed symmetrically around the trapping region. This facilitates keeping an ion near the RF-null while applying the necessary voltages for a crystal rotation: opposing pairs of electrodes with respect to the trapping region share the same voltages throughout a voltage sequence, thus keeping the field at the trapping region at zero. In 2D traps, however, the lower level of symmetry requires a finer understanding of the potentials generated by each electrode in order to avoid offsetting ions in the direction perpendicular to the trap surface during a voltage sequence. The trapping potential in this axis typically has the highest degree of anharmonicity, and the region of the lowest potential barrier for ion escape. Therefore, to avoid undesired potentials or ion loss, it is essential to keep ions near the RF-null of a surface trap.

\begin{figure*}
	\includegraphics[width=\textwidth]{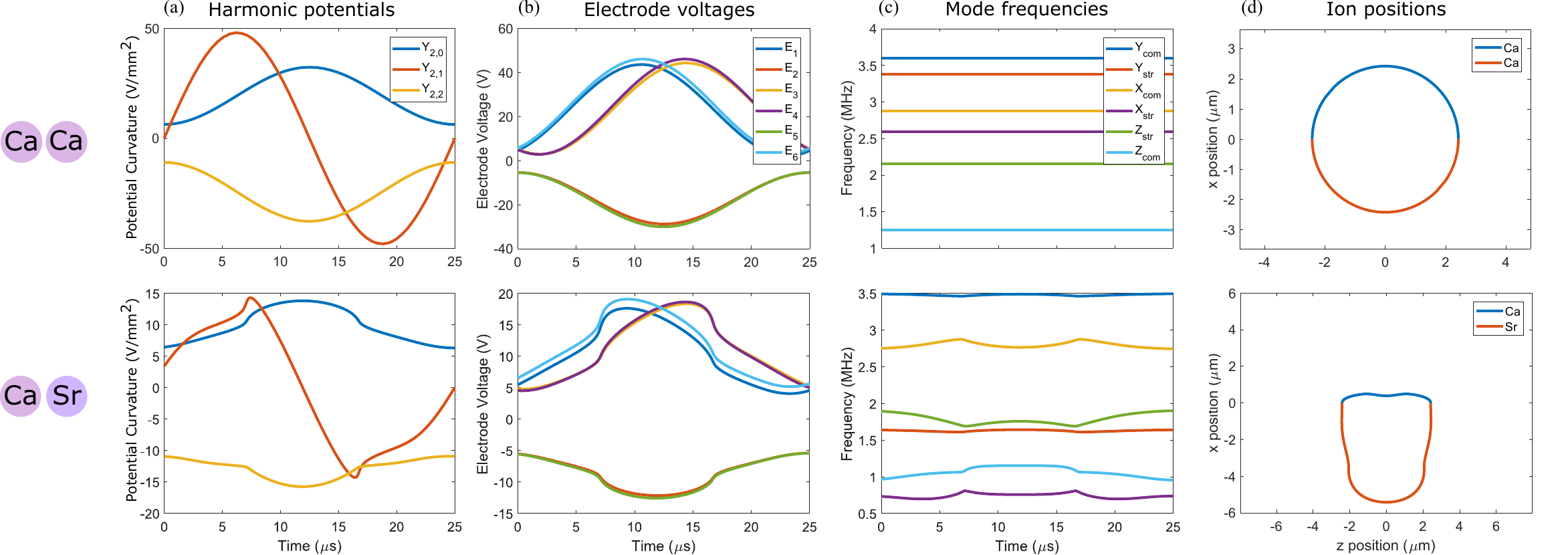}
	\caption{(a) Simulated second order potentials, where $Y_{2,0}$ represents axial confinement and $Y_{2,1}$ the diagonal offset in the $xz$-plane. The decrease in $Y_{2,2}$ lowers confinement in the $x$ direction. These potentials are optimized to minimize changes in trap frequency during the rotation. (b) Voltage sequence applied to nearby electrodes from the inset of Figure \ref{fig:Multipole_Potentials}(a). Further electrodes have been omitted, but behave similarly. Voltages on corner electrodes ($E_1$, $E_3$, $E_4$, and $E_6$) are increased, while those on central electrodes ($E_2$, $E_6$) are decreased, which results in higher axial confinement. The asymmetry in voltage on the corner electrodes provides a diagonal offset, enabling a rotation. Required voltages are based on simulations, and do not include simulation imperfections and effects of signal filtering. (c) Resulting trap frequencies, where the $z$ mode follows the axis of the ion crystal. Unlike that for a Ca - Ca crystal, Ca - Sr crystal mode frequencies cannot be kept constant throughout a rotation. (d) Ion positions during an optimized rotation. Due to the mass dependence of the pseudo-potential, Ca - Sr crystal rotations are asymmetric.}
    \label{fig:rotation_schematic}
\end{figure*}

Figure \ref{fig:rotation_schematic}(a) shows the sequence of spherical harmonic potentials (see Figure \ref{fig:Multipole_Potentials}(b) and Equation \ref{eq:spherical_harmonics}), as calculated using the method described in section \ref{sec:trap_potentials}, shown for both Ca - Ca and Ca - Sr crystals.
We vary $\theta_n$ from 0$^{\circ}$ to 180$^{\circ}$ in $n=100$ steps to ensure a smooth change in electrode voltages, since sudden changes in voltages contain frequency components that can excite the ion chain's motional modes.

The resulting waveform can be understood intuitively: $Y_{2,0}$ applies the axial confinement and radial anti-confinement that shifts the ion chain's axis from being along $z$ to being along $x$, effectively a 90$^\circ$ rotation. The potential given by $Y_{2,1}$ creates a diagonal offset in the $xz$ plane, ensuring that this reordering occurs in a specified direction. Once the ions are aligned along the $x$ axis, the confinement $Y_{2,0}$ is reduced, now with an opposite diagonal offset $Y_{2,1}$, completing a rotation. In order to facilitate rotations at lower voltages and increase control of the ions' motional frequencies, the confinement in the $x$ direction is lowered by decreasing $Y_{2,2}$.

Using the methods described in Section \ref{sec:trap_potentials}, we determine the voltages required to produce the potentials in Figure \ref{fig:rotation_schematic}(a). The voltages are displayed in (b), along with the associated motional frequencies (c), and ion trajectories (d), all shown for Ca - Ca (top) and Ca - Sr (bottom) rotations.

Multi-species rotations follow the same fundamental principle as described above, though come with added complications. Ions of different mass feel a different force when displaced from the RF null during a rotation because the curvature of the RF pseudo-potential is mass dependent ($\propto m^{-1}$). Thus, in contrast to a Ca - Ca crystal, the center of mass of a mixed-species crystal will be displaced from the RF-null throughout the rotation, unless a compensation is applied. The ions' positions during a rotation are displayed in Figure \ref{fig:rotation_schematic}(d), from which the asymmetry in a mixed species rotation is apparent. 

Also in contrast to single-species crystals, excess fields can be detrimental to the success of a mixed-species rotation. Due to the pseudo-potential's mass dependence, excess fields drive ions in a multi-species crystal away from the RF-null unequally.  Any field in the $x$ direction (radial, parallel to the trap plane) therefore results in an added diagonal offset in the $xz$ plane. In typical ion trapping potentials, an offset field on the order of $E_x=$ 0.01 V/mm produces a diagonal tilt similar to what would be caused by a potential curvature on the order of 1 V/mm$^2$ of the $Y_{2,1}$ spherical harmonic potential. This implies that voltage deviations of 10\% from their intended values can result in deviations of more than 20\% from the intended $Y_{2,1}$ potential. This deviation can disrupt successful bi-directional rotation, from which it is apparent why these potential terms need to be calibrated (see Section \ref{sec:potential_calibration}). If the $E_x$ field is well understood, an intentional offset in this field can be used advantageously in assisting uni-directional rotations, as previously demonstrated \cite{Home_NJP_2011,Chou_PRL_2010}. Up until now, mixed-species reordering has only been used uni-directionally as state preparation, and not as bi-directional swap for use a computation sequence. Minimizing heating, decoherence, and swap duration of mixed species ion crystals was therefore not crucial, and has not been considered previously.

Both single and mixed species crystal rotations thus require accurate control over a time-varying potential, generated by applying a sequence of voltages to the trap electrodes. The effect of electrical filters on these voltage sequences is discussed in the following section.

\subsection{Corrections for electrical filters}
\label{subsec:filters}
Voltage noise on trap electrodes presents a major contribution to mode heating of trapped ions \cite{Brownutt_RevModPhys_2015}, and needs to be minimized. Electrical filters attenuate voltage noise at ions' trap frequencies, and are therefore an essential component of an ion trap setup. In addition to suppressing electrical noise, filters will attenuate and slew an applied voltage sequence. Thus, the effect of filters needs to be taken into account when designing voltage sequences to realize trap potentials for a crystal rotation. In this section the employed filter system and the effect thereof on rotation sequences is discussed. 

\begin{figure}
	\includegraphics[width=\linewidth]{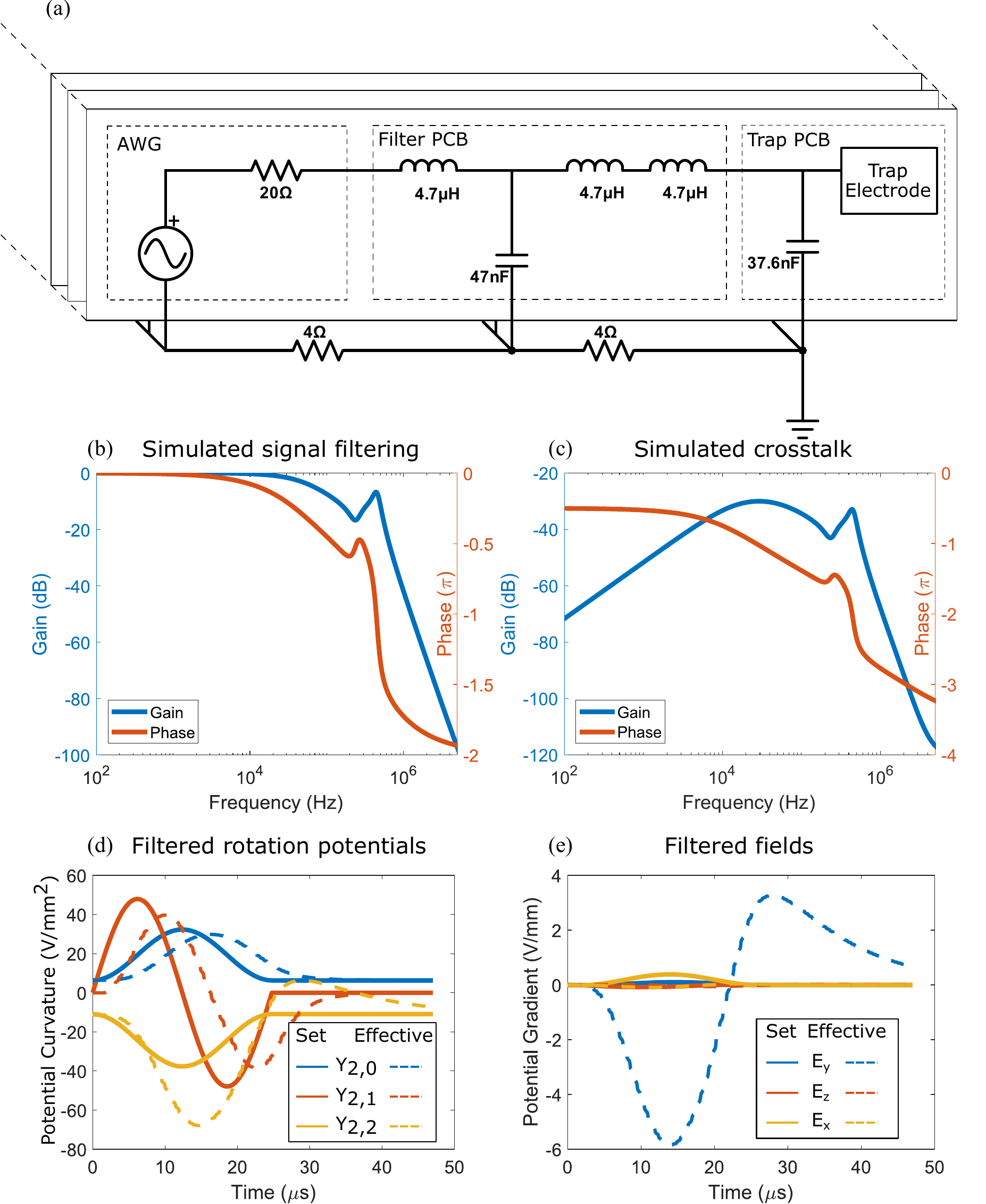}
	\caption{(a) Electronic schematic of the filtering network for a single electrode. (b) Simulated direct filter function at the trap electrode. (c) Simulated crosstalk, showing voltage pickup of other electrodes due to a voltage applied to a single electrode. (d) Set rotation sequence potentials, and simulated resulting potentials due to electrode filtering. (e) Effects of filtering on offset fields. Filter crosstalk has a significant effect on the field in the y-direction, leading to excessive micromotion.}
    \label{fig:filter_figure}
\end{figure}

Our filter setup for a single electrode is schematically shown in Figure \ref{fig:filter_figure}(a). The filter system consists of a two-stage LC low-pass filter (2 x 4.7$\mu$H and 47 nF), per electrode, incorporated in the cryostat's outer heat shield ($\sim$150 K). Additional filtering is performed in the vicinity of the trap with 8 parallel 4.7 nF capacitors. The cut-off frequency of 47 kHz has been chosen so that it sufficiently filters noise around the motional mode frequencies (42 dB at 1 MHz, and 80 dB at 3 MHz), while allowing voltage manipulation of the electrodes on time scales of tens of microseconds.
Inside the vacuum chamber are two connections to every electrode, each with separate but identical filter components. In practice, the second connection is only used to test the connectivity throughout the entire system, but cannot be neglected for filter analysis. The output amplifier of the AWG is included in the filter model as an effective series resistance (20 $\Omega$), which is taken as an estimate based on measuring the frequency response of the AWG directly at the output. 

Furthermore, the finite resistance of the electrical wiring is included in the filtering model. In a cryogenic system, electric wiring represents a major part of the heatload and thus there is a trade-off between low thermal conductivity and good electrical connectivity. We employ phosphor bronze cables, with a resistance of 4 $\Omega$/m at cryogenic temperatures. The ground connection for all electrodes consist of a twisted pair of these wires that  have a non-negligible resistance of about 4 $\Omega$, both from the vacuum feedthrough to the first filter stage and from the filter stage to the trap PCB. This finite wire resistance has the detrimental side-effect that the reference potential as supplied at the AWG output is not identical to the potential at the filters, or at the trap, if current is flowing through the capacitors.

The filter's transfer function at the trap, as determined by calculations based on ideal components, is shown in Figure \ref{fig:filter_figure}(b). Typical voltage sequences for rotations with a duration of 25 $\mu$s have their main frequency component at 20 - 40 kHz. Thus the applied electrode voltages are attenuated by 1.5 - 2 dB, and their waveforms experience a delay of about 5 $\mu$s. Figure \ref{fig:filter_figure}(d) shows how the waveforms of the significant rotation potentials $Y_{2,\lbrace0,1,2\rbrace}$ are affected by filtering.
The distortion in the potential waveforms can impair the performance of a rotation or even impede rotations altogether. 

As a result of the finite ground wire resistance, supplying a time-dependent voltage to one electrode will change the reference potential at the trap. 
The voltage difference between AWG and trap ground levels results in an effective crosstalk of electrodes. The simulated crosstalk transfer function between any two electrodes is displayed in Figure \ref{fig:filter_figure}(c). Upon applying a typical voltage sequence to a single electrode, we measure a crosstalk voltage of at most 3\% on all other electrodes, with a settling time of about 20 $\mu$s. However, a full rotation sequence employs all electrodes, with a combination of positive and negative voltages. During a typical rotation, the effective crosstalk voltage is between 0.5 and 1.0 V, resulting in a voltage error between 3\% and 10\% for different electrodes. This voltage error perturbs the intended potentials, most notably adding an undesired electric field perpendicular to the trap plane ($Y_{1,-1}$) of up to 6 V/mm, as can be seen in Figure \ref{fig:filter_figure}(e).

The inclusion of this undesired field due to filtering highlights the complexity of physical ion crystal manipulation on surface traps, compared to 3D traps. In 3D traps, opposing electrodes maintain similar voltages throughout a voltage sequence, even when filtered, and thus ion crystals will remain near the RF-null due to symmetry. The geometry of surface traps does not have symmetry around the trapping region perpendicular to the trap plane. Therefore, if a calculated voltage sequence is designed to keep micromotion compensated throughout, the \textit{filtered} sequence will generate an additional electric field perpendicular to the trap plane.

The unintended offset field pushes the ions further from the RF null, thus substantially increasing micromotion. Micromotion is not necessarily detrimental to an ion's motional state: experiments where we intentionally induce small excursions of micromotion have resulted in negligible heating of less than 0.1 phonons after being subjected to a field of $\pm$0.5 V/mm for 200 $\mu$s. However, at the simulated 6 V/mm of offset field, the ions do not remain crystallized, let alone remain near the motional ground state. For some of the calculated voltage sequences, the unwanted field is sufficient to expel ions from the trap, since along the perpendicular axis the escape energy is typically the lowest. Therefore, a method of compensating electrode voltage sequences in order to achieve low-heating rotations is required. Through simulations, we deduce which filtered voltage sequence is required in order to achieve the \textit{intended} voltage sequence at the trap electrodes. 

The transfer functions displayed in Figure \ref{fig:filter_figure}(b) and (c) are based on ideal electrical components and environment. In practice, tolerances in components, and other sources of inductance, capacitance, and crosstalk, causes the modelled transfer function to deviate from the physical electrical response. 
The model thus represents an approximation that is necessary for creating filter-corrected voltage sequences for ion crystal rotations. We emphasise that without the application of filter and potential calibration models to our solutions, our voltage solutions typically do not lead to successful rotations at the desired time scales (20 - 30 $\mu$s). Voltage solutions that take our models into account lead to successful rotations that, despite inducing a high level of motional heating, can be used as a starting point for further optimization. Our method of minimizing rotation-induced heating is outlined in the following section.

\subsection{Automated calibration}
\label{subsec:iter_filter_improve}
We aim to implement crystal rotations that cause the least motional mode heating. As a means to this end, we seek a set of electrode voltage waveforms at the trap electrodes that keep all motional mode frequencies of the ion crystal constant, to avoid low frequencies and frequency crossings. 
The multipole calibration in Section \ref{sec:potential_calibration} and filter model in Section \ref{subsec:filters} are sufficient to generate voltage sequences that produce rotations. However, to minimize heating, additional optimization must be performed.
We therefore utilize an automated optimization routine. There exist many possible optimization candidates, (hill climbing, direct stochastic, simulated annealing, genetic algorithms, etc.), each with advantages and disadvantages. We use a machine-learning approach, based on a single hidden layer neural network \cite{Murphy_2012} (using the MATLAB\textsuperscript\textregistered~Neural Network Toolbox\texttrademark), to find the set of waveforms that has the least influence on the motional state of the ion crystal. We choose this method for its relative computational simplicity, its ability to explore a high dimensional parameter space, and resilience to finding solely local optima. Furthermore, neural networks have the advantage that one does not require a predetermined model that describes the relation between input (sequence parameters) and output (motional heating). 

\begin{figure}
	\includegraphics[width=\linewidth]{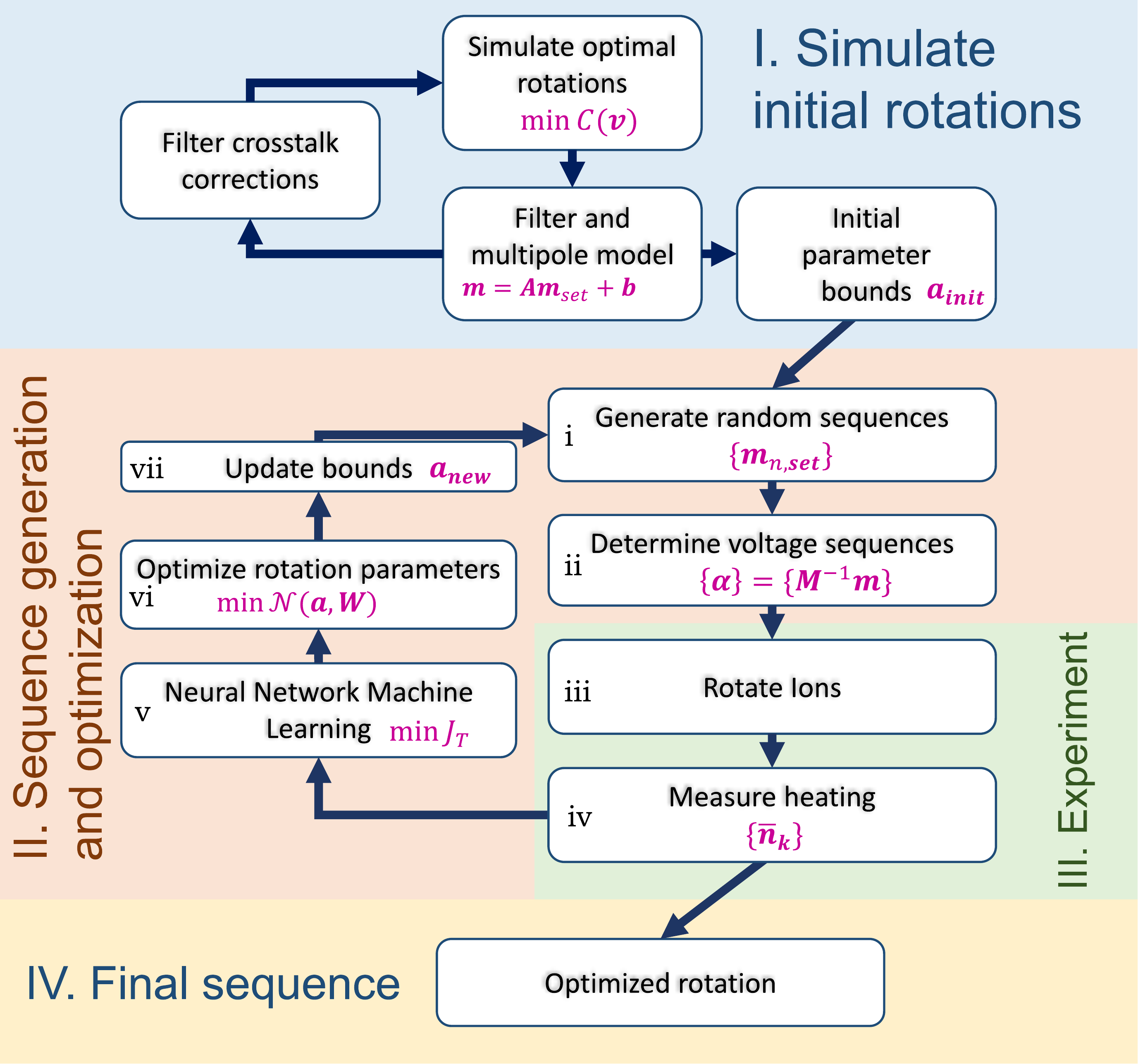}
	\caption{Sketch of the procedure to obtain the optimized rotation parameters. (I) An initial guess for sequence parameters is made by simulating ideal rotations, and iteratively applying our filter model and multipole calibrations to optimize motional frequencies and minimize micromotion. (II) Sequence parameters are optimized using a machine learning neural network, which is trained using (III) experimentally determined rotation induced heating. (IV) An optimized rotation sequence is used in further experiments. The symbols in red indicate parameters relevant to that step, of which the description can be found the main text.}
    \label{fig:learning_diagram}
\end{figure}

Figure \ref{fig:learning_diagram} schematically describes our method of achieving optimized ion crystal rotations, which is based on the following four steps: (I) An initial parametrized rotation sequence is simulated, using the calibrated multipole and filter models. (II) Using this sequence as a starting point, a neural network based learning algorithm is used to optimize rotation sequences to minimize induced motional heating. (III) Experimental data, where heating is measured for various rotation sequences, is used to train the network, from which an optimized rotation is found. (IV) This optimized rotation is used in further experiments. These steps are further detailed below. 

\textbf{Step I:} For the optimization, we first parameterize the sequence of applied potentials obtained using the procedure described in Section \ref{sec:trap_potentials}.
Figure \ref{fig:rotation_schematic}(a) shows the trapping potentials that lead to a rotation with constant mode frequencies. We choose a Fourier component parametrization, since the simulated waveform of the potentials in Figure \ref{fig:rotation_schematic}(a) can be approximated by a few orders of a Fourier expansion. We can express the sequence of applied potentials, discretized into $N$ discrete steps, as
\begin{equation}
\label{eq:potential_parametrization}
\textbf{m}_{n,set}=\textbf{a}_0 + \sum_{p=1}^{p_{max}} \textbf{a}_p \sin\left(p \pi n/N\right).
\end{equation}
Here $\textbf{m}_{n,set}$ is a vector of spherical harmonic potentials (all 8 terms of degrees $l=1$ and $l=2$), that characterize the set potential at step $n$. $\textbf{a}_p$ is a set of weights that describe the amplitudes of the Fourier components of the potentials. $p_{max}$ is a chosen maximum order of the Fourier components. The higher order Fourier components in the waveforms ($p>1$) can approximately compensate for the distortion that the filters induce to the waveform. For our optimization approach, we choose $p_{max}=2$ to limit the number of optimization parameters $\textbf{a}_p$, thus reducing the number of parameters in the neural network used for optimization.

In addition to varying the total rotation time (by adjusting $N$), we have 24 weights that parametrize the sequence ($\textbf{a}_{\left\lbrace0,1,2\right\rbrace}$, each with 8 terms), giving a total of 25 parametrization coefficients.  It is not required to include the sample rate of the sequence (2 MHz) as an optimizable sequence parameter, as this frequency is heavily filtered in our set-up. For notational simplicity, we cast all parametrization coefficients into a single vector, $\textbf{a}$. 

We aim to generate the potential sequence of Figure \ref{fig:rotation_schematic}(a), using parametrization coefficients $\textbf{a}$. The desired sequence $\textbf{m}_{n,des}$ differs from the set set potential $\textbf{m}_{n,set}$, due to static offsets described by our potential calibration model (Section \ref{sec:potential_calibration}) and dynamic offsets described by our filter model (Section \ref{subsec:filters}).  We numerically find a set of coefficients $\textbf{a}_{init}$ that, under the influence of these offsets, result in the intended potentials $\textbf{m}_{n,des}$.

\textbf{Steps II - III:} The sequence of voltages given by the multipole potentials $\textbf{m}_{n,set}$ at $\textbf{a}_{init}$ results in tens of quanta of rotation induced heating. The coefficients in $\textbf{a}_{init}$ therefore need to be adjusted to minimize heating. We focus on minimizing heating on the axial modes, but the described method is applicable to any set of motional modes. We do not adjust the 0$^{\textrm{th}}$-order Fourier components of the parametrized sequences ($\textbf{a}_0$ in Eq. \ref{eq:potential_parametrization}), which govern the static potential before and after a rotation, to avoid possible changes in mode cooling before a sequence.

The simulated sequence parametrized by $\textbf{a}_{init}$ is used as an initial point for the optimization procedure. In our procedure, the following steps are taken: (i) a large amount of sequences with parameters that marginally deviate from $\textbf{a}_{init}$ are generated. (ii) Electrode voltages that realize these sequences are determined using the method in appendix \ref{sec:appendix_voltagesolutions}. (iii) The voltage sequences are applied in the experiment to rotate an ion crystal, from which (iv) heating is measured. (v) The heating results are used as targets to train a neural network, which has the rotation sequence parameters \textbf{a} as inputs. (vi) The trained network is used to find a new parameter set $\textbf{a}_{new}$ that, according to the network, should result in minimal heating. (vii) This new parameter set is used as a new central point around which to create a set of sequences, as in (i). Steps (i) - (vii) constitute a single iteration of the optimization. These steps are repeated several times, until the measured heating rate converges. We will outline the optimization procedure in more detail below. 

\textbf{(i)} In our experiment, for the initial training iteration, $K=500$ rotation sequences are generated by randomly selecting a set of weights $\lbrace\textbf{a}_k\rbrace$ within given bounds around the initial parameter set $\textbf{a}_{init}$. We have chosen bounds by experimentally finding limits on parameters that do not result in expulsion of ions from the trap, and result in rotations. Bounds that are within a safe limit of these requirements are typically $\pm$2 - 5\% of the respective coefficients of $\textbf{a}_{init}$. \textbf{(ii)} The voltages $\bm{\alpha}$ required to realize this set of rotation sequences are determined with $\bm{\alpha}_n=\textbf{M}^{-1}\textbf{m}_{n,set}$ (see appendix \ref{sec:appendix_voltagesolutions}), where $\textbf{m}_{n,set}$ is given by Equation \ref{eq:potential_parametrization}. \textbf{(iii)} These voltage sequences are applied to an ion crystal, and \textbf{(iv)} the resulting motional heating is measured. The non-optimized sequences add 10 - 100  quanta to the motional state. In this regime, the added energy can be estimated by the decay of Rabi oscillations when resonantly driving a carrier transition. For motional heating in the range of less than 10 quanta, we monitor the excitation of the motional sidebands \cite{Diedrich_RhysRevLett_1989}. The resulting excitation curves are fitted to an analytical model, derived from the two-ion interaction Hamiltonian given by the Jaynes-Cummings model (see Appendix \ref{sec:appendix_JC}) to find the excess energy resulting from a rotation, in terms of mean phonon numbers $\bar{n}$. The mean phonon numbers are used as a target to train a neural network with the sequence parameters $\textbf{a}$ as features.

\textbf{(v)} We use a neural network, denoted by $\mathcal{N}$, that takes sequence parameters $\textbf{a}$ as an input (the neural network's features \cite{Murphy_2012}) to make a prediction of the resulting mode heating $\bar{n}_{pred}$ (the target). The network function is thus given by $\bar{n}_{pred}=\mathcal{N}(\textbf{a},\textbf{W})$ where $\textbf{W}$ are weights and biases of the network. The network must first be trained with experimental data in order to make realistic predictions.

Training is realized by supplying the network with experimental data, consisting of the set of rotation sequence parameters $\lbrace\textbf{a}_k\rbrace$ and experimentally determined resulting mode heating $\lbrace\bar{n}_k\rbrace$. The network is trained by minimizing the network cost function $J_T$, which describes the deviation between the network's predictions for the set of sequence parameters and the real measured values. The network error function for a single training measurement, $J_k$, consisting of parameters $\textbf{a}_k$ and measured mean phonon number $\bar{n}_k$ is given by $J_k=\left(\bar{n}_k-\mathcal{N}(\textbf{a}_k,\textbf{W})\right)^2$. The total cost function is then $J_T = \sum_k J_k$.
The network weights and biases $\textbf{W}$ are optimized by finding the minimum of the cost function $\min_{\textbf{W}}(J_T)$. 

\textbf{(vi)} The trained network is used to find a parameter set $\textbf{a}$ that minimizes the \textit{predicted} mean phonon number $\bar{n}$, as $\min_{\textbf{a}}\left(\mathcal{N}(\textbf{a},\textbf{W})\right)$, where $\textbf{W}$ are the trained set of weights and biases of step (v). As the network has been trained for parameters $\lbrace\textbf{a}_k\rbrace$ within specific bounds, the parameter set that minimizes $\mathcal{N}(\textbf{a},\textbf{W})$ should be constrained to the same bounds.

The network $\mathcal{N}$ typically has multiple local minima. Therefore, the calculated value of $\min_{\textbf{a}}\left(\mathcal{N}(\textbf{a},\textbf{W})\right)$ depends on the supplied initial values of $\textbf{a}$. We use parameters corresponding to the best results from the scan (typically the lowest 10\% mean phonon numbers $\bar{n}$) as starting points for the minimization. For our experimental data, this post-selection results in a unanimous consensus for the value of $\textbf{a}_{min}$ that minimizes $\mathcal{N}$. This parameter set is used as a central point around which new random sequences are generated. Step (i) is thus repeated, with a new central value $\textbf{a}_{min}$.

Steps II and III are iteratively repeated, until the measured heating rate no longer decreases. In our experiment, we have seen that after four iterations, the optimization routine generally converges.

\textbf{Step IV:} The final iteration produces another series of randomly generated rotations, with a mean heating rate of a few phonons. From this set, the result with the lowest measured heating is selected. Section \ref{sec:exp_results} discusses further experimental results which utilize an optimized rotation sequence.

%% file: results.tex
\subsection{Single species rotations}
We present the results for rotations in a crystal consisting of two Ca ions. Initially (ie. with no rotation), trap frequencies are $\{\omega_x,\omega_y,\omega_z\}=2\pi\cdot\{ 2.796(1), 3.499(1), 1.257(1)\}$ MHz. At the beginning of an experimental sequence, we prepare both ions in the $S_{1/2}(m=-1/2)$ (ground) state. We then apply a global $\pi / 2$ pulse on resonance with the $4S_{1/2} \leftrightarrow 3D_{5/2}$ transition (at 729 nm), defined as 
\begin{equation}
\label{eq:pi_over_two}
R_{x}\left(\frac{\pi}{2}\right)=\exp(-i\pi/4(\sigma_x^{(1)}+\sigma_x^{(2)})),
\end{equation}
with Pauli operators $\sigma_x^{(i)}$ for ions 1 and 2 \cite{Nielsen_2011}.
This operation prepares the ions in the $1/2\left( \left|SS\right> + \left|SD\right> +\left|DS\right> +\left|DD\right>\right)$ state. State detection is performed by monitoring state-dependent fluorescence ($4S_{1/2}\leftrightarrow 4P_{1/2}$, 397 nm) with a CCD camera \cite{Schindler_NJP_2013}. Detection is performed before and after a rotation. A successful rotation is indicated by a switch from $\left|SD\right>$ to $\left|DS\right>$ and vice versa.
Table \ref{tab:truth_table} displays the resulting truth table. The data for each number of rotations consists of 250 repetitions of the experiment which are post-selected to include only the $\left|SD\right>$ and $\left|DS\right>$ initial states. For all numbers of rotations, including no rotations, a wait time of 250 $\mu$s is used between detection events, which is enough time to allow all rotations to finish. After one rotation, 94(3)\% of initial states of $\left|SD\right>$ are measured to end up in  $\left|DS\right>$, and 90(5)\% vice versa. The deviations from 100\% are attributed to limitations caused by the CCD camera's readout noise in ion state detection. Readout errors are caused by spontaneous decay to the $S$ state during the 50 ms detection time (corresponding to a 4\% decay probability after the first detection, and 8\% after the second detection). As can be seen in Table \ref{tab:truth_table}, 0\% of cases in which states are initially in $\left|SD\right>$ or $\left|DS\right>$ remain unchanged after one rotation. Therefore, despite detection errors, we can conclude that the success probability of the rotation is 100\%$_{-1}$. The same conclusion can be drawn for multiple rotations in succession, as the success probability does not depend on the number of rotations.

\input{table}

\begin{figure}
	\includegraphics[width=\linewidth]{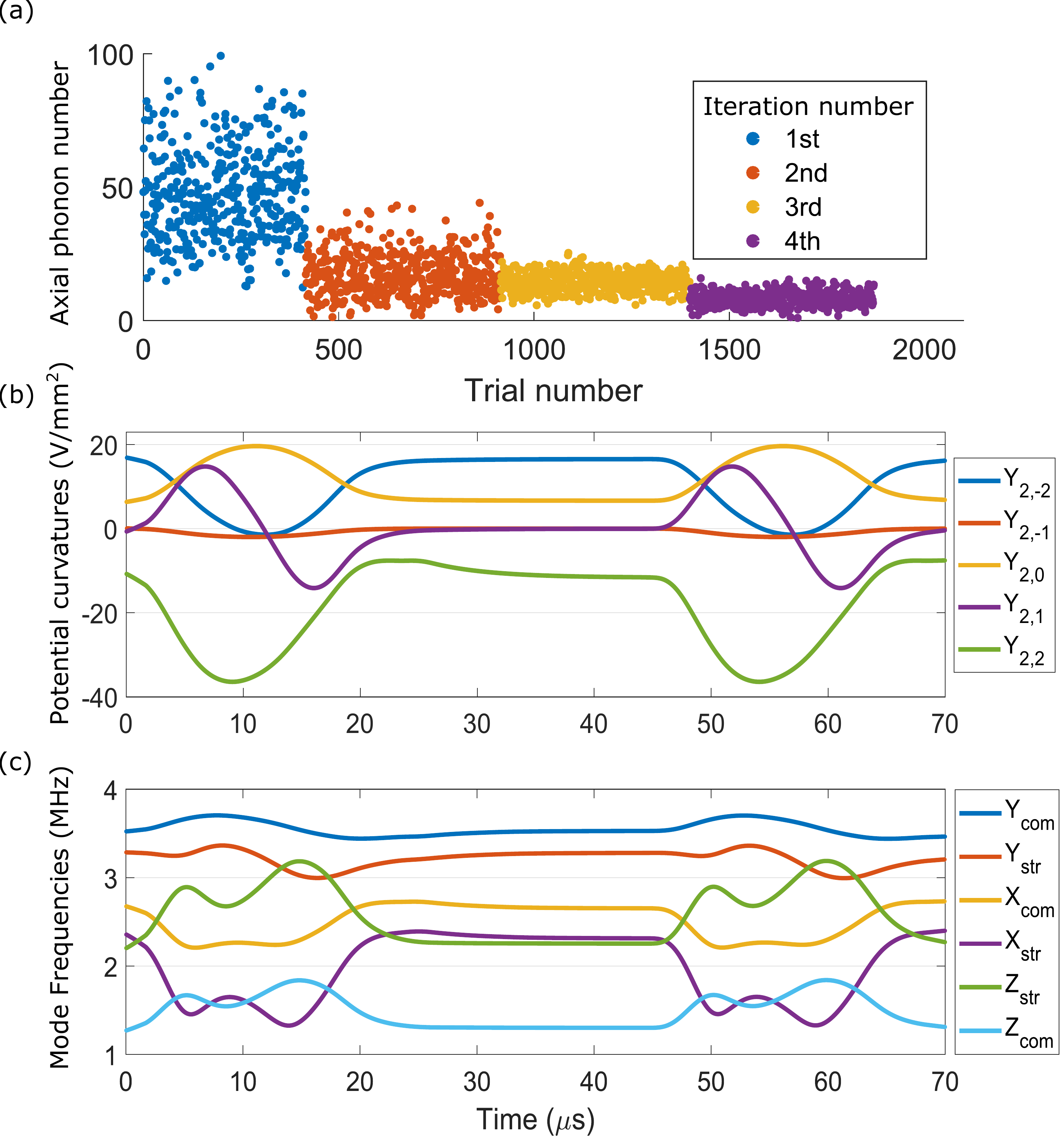}
	\caption{(a) Measured rotation induced heating during four iterations of the automated sequence optimization. We simulated the harmonic potentials of the the selected rotation sequence (b) and their associated mode frequencies (c), shown for two consecutive rotations with 20 $\mu$s wait time between them. The $Z$-modes rotate along with axis of the ion chain, and the $Y$-modes are the axis of rotation. Included in the simulation are potential offsets (Section \ref{sec:potential_calibration}) and voltage filtering (Section \ref{subsec:filters})}
    \label{fig:iterative_improvement}
\end{figure}

We employ the method described in Section \ref{subsec:iter_filter_improve} to minimize rotation-induced heating. The total number of steps in a sequence is included as an optimization parameter, and converges around 50 steps with a fixed step size of 500 ns, giving a total rotation time of 25 $\mu$s. 
Figure \ref{fig:iterative_improvement}(a) displays mean phonon numbers as measured in the first four iterations of the optimization procedure. Measuring phonon numbers does not require state detection of individual ions, merely that of the collective chain. We therefore opt to use a PMT instead of a CCD camera for state detection in all heating measurements, which reduces detection time in our setup by an order of magnitude.

Four iterations of the optimization procedure are required to improve the non-optimized calculated rotation sequence, with tens of rotation-induced phonons, to sequences with single phonons. After four iterations the optimization procedure has converged and no longer decreases heating.

The sequence with the lowest heating rate from the final optimization iteration is used in further experiments. The simulated spherical harmonic terms of the potential and mode frequencies of this sequence are plotted in Figures \ref{fig:iterative_improvement}(b) and (c). The displayed potentials are obtained by passing the applied voltage sequence through the filter simulation, and expanding the resulting potential in terms of the spherical harmonics. The mode frequencies are calculated from these potentials. In Figure \ref{fig:iterative_improvement}(c) it can be seen that the frequencies of several motional modes cross during a rotation sequence. Notably, the $X_{str}$ and $Z$ modes cross several times. While mode frequencies are near degeneracy, exchanges of phonons are more likely \cite{Marquet_ApplPhysB_2003}. Experimental results are therefore sensitive to the initial phonon number of $X_{str}$. We thus intentionally apply a potential of $Y_{2,-2}=16$ V/mm$^2$ to increase the coupling of the Doppler-cooling beam to the $X$ modes, resulting in a lower initial phonon number for the $X$ modes. In separate experimental trials from those described above, we have seen that this initial tilt reduces rotation-induced axial stretch mode heating by more than a factor of two.

\begin{figure*}
	\includegraphics[width=\linewidth]{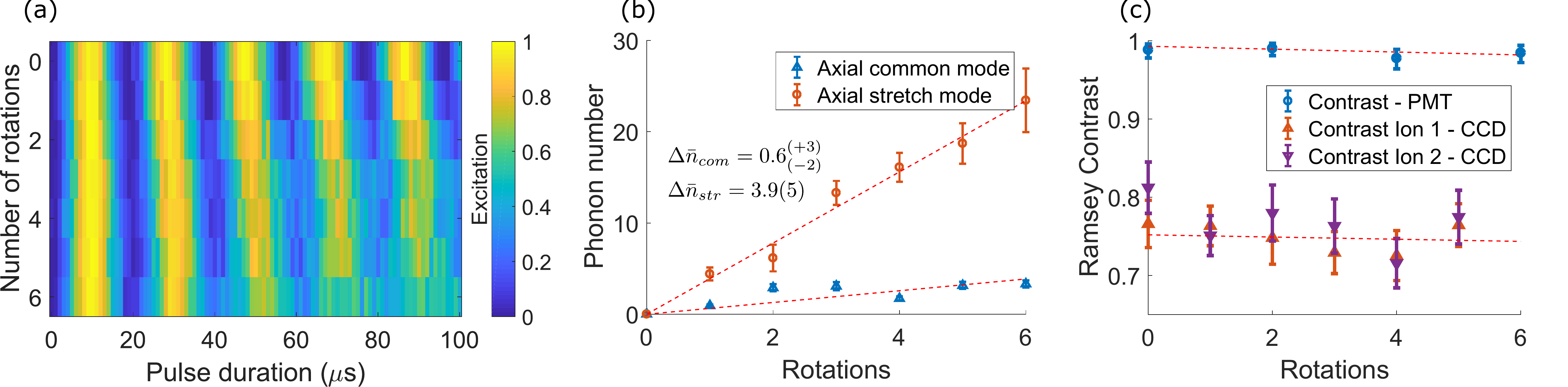}
	\caption{(a) Rabi oscillations on a carrier transition and (b) measured mean phonon numbers on the axial common and stretch modes after a number of rotations. (c) Qubit coherence after a number of rotations. The dashed red lines are least-squares fits, with linear model in (b) and an exponential decay model in (c)}
    \label{fig:rotation_results}
\end{figure*}

Before a rotation, the axial modes are sideband-cooled to an average phonon number of the common mode $\bar{n}_{com}=0.017(16)$ and the stretch mode $\bar{n}_{str}=0.011^{(+4)}_{(-5)}$. After various numbers of rotations, we excite the $S_{1/2}\leftrightarrow D_{5/2}$ carrier transition to obtain Rabi oscillations, displayed in Figure \ref{fig:rotation_results}(a). Figure \ref{fig:rotation_results}(b) shows the mean phonon numbers for both modes as a function of the number of rotations, as estimated using sideband excitation data. From these results we determine the axial heating rates to be $\Delta \bar{n}_{com}=0.6^{(+3)}_{(-2)}$ phonons per rotation on the common mode, and $\Delta \bar{n}_{stretch}=3.9(5)$ on the stretch mode. Stationary heating rates have been determined to be several phonons per second in our trap \cite{Brandl_RSI_2016} and are therefore negligible with respect to the rotation-induced heating rates.

The decay at a higher number of rotations in Figure~\ref{fig:rotation_results}(a) is attributed to, and consistent with, the measured axial heating in Figure~\ref{fig:rotation_results}(b). Motional heating in the axial direction results in an excitation of a carrier transition given by:
\begin{equation}
\sum_{n_{com}=0}^\infty \sum_{n_{str}=0}^\infty P_{\bar{n}}(n_{com}) P_{\bar{n}}(n_{str}) \sin^2(\Omega_{n_{com},n_{str}} t/2)
\end{equation}
where $P_{\bar{n}}(n)$ is the occupation probability to have $n$ phonons, given a mean phonon number $\bar{n}$. The occupation probability for a thermal distribution is given by
\begin{equation}
P_{\bar{n}}(n) = \frac{1}{\bar{n}+1}\left(\frac{\bar{n}}{\bar{n}+1}\right)^n.
\end{equation}
$\Omega_{n_{com},n_{str}}$ is a coupling constant (in the second order Lamb-Dicke approximation):
\begin{equation}
\label{eq:coupling_constants}
\Omega_{n_{com},n_{str}} = \Omega_{0,0}\left(1-\eta_{com}^2 n_{com} - \eta_{str}^2 n_{str}\right),
\end{equation}
with $\Omega_{0,0}$ the ground state carrier coupling constant, and $\eta_{com}$ and $\eta_{str}$ the mode-frequency dependent Lamb-Dicke factors, 0.061 and 0.047 in our particular setup.

We attribute the heating in the stretch mode to phonon exchange during moments of degeneracy with the $X_{str}$ and $Y_{str}$ modes, as can be seen in Figure~\ref{fig:iterative_improvement}(c). While a local parameter optimum has been found with our optimization procedure, the resulting evolution of mode frequencies in Figure~\ref{fig:iterative_improvement}(c) suggests that a more stable solution, in which mode frequencies do not cross, might exist.

Our efforts in reducing heating caused by ion crystal rotations stems from the requirement to minimize loss of qubit coherence, imperative when considering rotations for use as coherent operations in quantum computational sequences as a SWAP gate \cite{Tan_Nature_2015}. We have demonstrated that our minimization routine results in rotation sequences that induce less than 1 phonon on the common axial mode. In the following, we show that qubit coherence is maintained after applying an optimized rotation sequence. 

To assess how well coherence is preserved, we first prepare the ions in the same state as for measuring rotation success probability (Equation \ref{eq:pi_over_two}). Then, after a set of rotations, coherence is measured by applying an analysis pulse, given by
\begin{equation}
R_{\phi^{(1)},\phi^{(2)}}\left(\frac{\pi}{2}\right)=\exp(-i\pi/4(\sigma_{\phi^{(1)}}^{(1)}+\sigma_{\phi^{(2)}}^{(2)})),
\end{equation}
with $\sigma_{\phi^{(i)}}^{(i)}=\cos(\phi^{(i)})\sigma_x^{(i)}+\sin(\phi^{(i)})\sigma_y^{(i)}$. Here, $\phi^{(i)}$ is a phase shift experienced by ion i between the initial and analysis pulse. This phase consists of a controlled phase change in the light field, $\phi_{var}$, a phase evolution due to experimental drifts and fluctuations, $\phi_{drift}$ (ex. magnetic field and laser frequency), and a phase difference due to changes in the ions' position $\phi^{(i)}_{pos}$: 
\begin{equation}
\phi^{(i)}=\phi_{var}+\phi_{drift}+\phi^{(i)}_{pos}
\end{equation}
Scanning $\phi_{var}$ produces Ramsey fringes, where the fringe contrast is an indication of qubit coherence. The resulting contrast after multiple rotations is shown in Figure \ref{fig:rotation_results}(c). For every number of rotations, including zero, the total wait time between the first and second Ramsey pulse is kept at 300 $\mu$s, so that ions experience the same dephasing caused by $\phi_{drift}$, independent of the amount of rotations. The time of 300 $\mu$s ensures a sufficient settling time for filtered voltages and thus ion positions, also for six rotations. The contrast of ~79\% at 0 rotations is attributed to CCD detection errors. Regardless of this initial contrast, we can infer loss of coherence from the decay in contrast for multiple rotations. Complementarily, we repeat the experiment with a PMT, which reduces detection errors to 0.2\%. The PMT does not differentiate between the $\left|SD\right>$ and $\left|DS\right>$ states, thus a contrast measurement with the PMT does not account for non-identical phase differences for the two ions, and will in general result in loss of contrast if $\phi^{(1)}\neq\phi^{(2)}$.  Since swapping the position of two ions generates an equal but opposite phase shift on both ions ($\phi_{pos}^{(1)}=-\phi_{pos}^{(2)}$), and therefore causes a loss in contrast, we omit data for odd numbers of rotations. The contrast of $98.8^{(+0.7)}_{(-1.0)}\%$ for 0 rotations is consistent with previously measured Ramsey contrasts of our system, at 300 $\mu$s. After applying 6 rotations, a contrast of $98.4^{(+0.9)}_{(-1.2)}\% $ is maintained. From a weighted exponential decay fit, using both the CCD and PMT data, we infer a coherence loss of 0.2(2)\% per rotation.

\subsection{Mixed species rotations}
Mixed species rotations use the same procedure as for single species rotations, described in Section \ref{sec:trap_potentials}. However, ensuring that micromotion remains well compensated throughout the rotation is more essential: since the RF pseudo-potential is mass dependent, ions are forced away from the RF null unequally under the influence of a stray field. An electric field in a radial direction will force Strontium ions $m_{Sr}/m_{Ca}=2.2$ further away from the RF-null than Ca ions. Consequently, controlling the position of Sr ions requires a higher precision in applied voltages than required for Ca. 
On the other hand, the Strontium ion's lower radial potential means that aligning Ca - Sr crystals along the x-axis can be achieved using a lower axial confining potential. Hence, successful Ca - Sr rotations require lower electrode voltages compared to those of Ca - Ca crystals, as can be seen in Figure \ref{fig:rotation_schematic}(b).

Applying the protocol described in Section \ref{sec:Rotation_potentials} (the same as for Ca - Ca crystals, though with different masses in equation \ref{eq:trap_potential}), we find suitable sequences of potentials that provide successful rotations. Determining the success rate of a rotation does not require detection of the internal states of the ions since both species fluoresce at different wavelengths. It is thus sufficient to monitor the position of the Ca ion. We opted to detect Calcium fluorescence with a PMT to obtain the data, since in our setup our PMT's detection rate is much higher than that of the CCD camera. We discriminate the two possible positions of the Ca ion by placing a knife edge in an intermediate image plane of the detection optics, resulting in a different count rate depending on the position of the Ca ion. We thus use a PMT count rate threshold to distinguish ion positions. In our setup, the two possible ion positions produce count rates that are $3\sigma$ from the threshold, resulting in less than 0.2\% detection error between the two states. The position is determined both before and after a rotation sequence. We have detected successful rotations for all of 100 rotation sequences, from which we infer a 100$_{(-1)}$\% success rate for Ca - Sr rotations.

\begin{figure}
	\includegraphics[width=\linewidth]{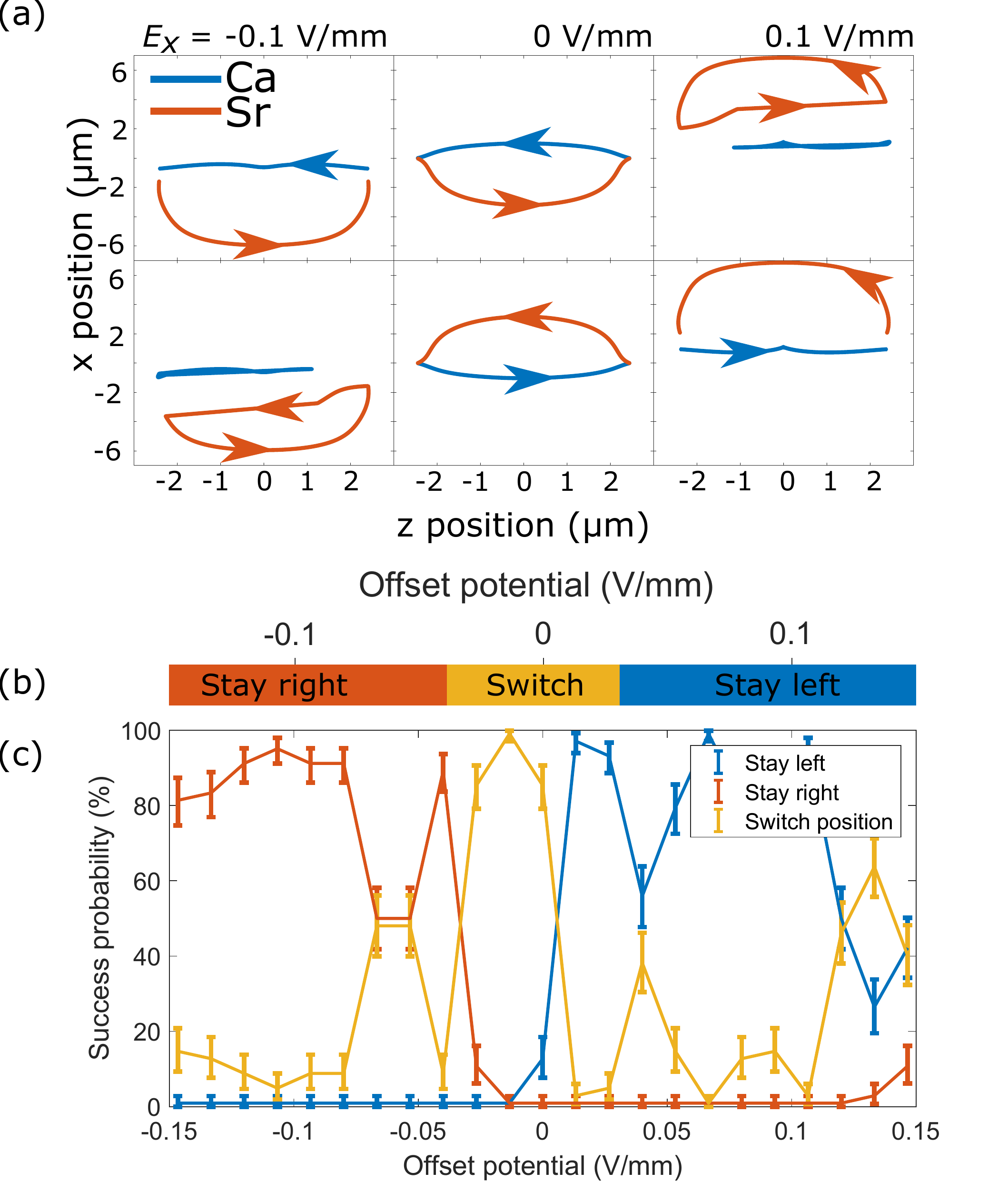}
	\caption{Ca - Sr crystal rotations under influence of a uniform external potential $E_x$ along the $x$-axis. (a) Simulated ion trajectories for various potentials $E_x$. The Sr ion starts on the left(right) side in the top(bottom) panels. The bottom left and top right trajectories show both ions ending in their original positions, thus indicating unsuccessful rotations. (b) Simulated rotation outcome for varying potentials, which depicts good overlap with (c) experimentally measured rotation success probabilities in varying offset fields. The experimental results show the possible outcomes of a Strontium ion in a mixed species crystal.}
    \label{fig:CaSrOffset}
\end{figure}

As described in Section \ref{sec:rotation_principle}, the success of mixed-species rotations is inherently more sensitive to stray fields, notably in the radial direction in the plane of rotation. We intentionally apply an offset field during a rotation to characterize this sensitivity. Figure \ref{fig:CaSrOffset} shows both simulated and measured rotation success, under the influence of different field strengths.
For the simulations, ion positions are calculated as function of applied potentials to determine whether a rotation is successful or not. Examples of simulated ion trajectories for various field strengths are shown in Figure \ref{fig:CaSrOffset}(a), where the left and right panels indicate a successful rotation in only one direction. Figure \ref{fig:CaSrOffset}(b) shows which applied offset potentials result in which regime according to simulations, and shows good agreement with measured data, shown in Figure \ref{fig:CaSrOffset}(c). 
In both the simulations and measurements, three distinct regimes are apparent: regardless of the initial configuration, a rotation with a negative offset field consistently results in one configuration (Ca - Sr), a positive field in the other (Sr - Ca), and a well-compensated field results in swapping back and forth between the two. 
For a reliable bi-directional rotation, the horizontal bias field should be compensated with an accuracy of less than 25 mV/mm.

The rotations' sensitivity to an offset field can be used to perform deterministic reordering of multi-species crystals, in a specified direction. This concept has previously been applied to multi-species ion chains, but only as state preparation in case of an accidental switch of ion position \cite{Hume_PhysRevA_2009}. Here we show that such operations can also be applied during an experimental sequence, prospectively with low loss of coherence for use as part of a quantum computation sequence.

After applying a rotation sequence to Ca - Sr crystals, we no longer resolve contrast in Rabi oscillations when exciting a carrier transition. From this, we estimate a combined mode heating of more than 80 quanta in the axial common and stretch modes. We have attempted applying the same optimization routine as used for Ca - Ca crystals for lowering heating. However, even with conservative bounds on optimization parameters, the routine frequently results in expulsion of ions from the trap, and does not display a significant reduction in heating. We attribute our filter set-up as a probable cause of rotation induced heating, both because it hinders us from exactly controlling motional frequencies during a sequence, as well as inducing excessive micromotion. Additionally, parameterizing our sequences to order $m_{max}$ can potentially result in sequences that do not sufficiently compensate offsets caused by filtering.

%% file: table.tex
\begin{table}
\centering
\caption{Truth table for state detection before and after a number of rotations. Values are the percentages that we detect a final state, for given an input state. For each number of rotations, 250 trails were performed, of which the initial states $\left|SD\right>$ and $\left|DS\right>$ were post-selected}
\label{tab:truth_table}
\begin{tabular}{llllll}
                                                                &                                                                              & \multicolumn{4}{c}{Final state}                                                                                                                                   \\ \cline{3-6} 
\begin{tabular}[c]{@{}l@{}}Number\\ of\\ Rotations\end{tabular} & \multicolumn{1}{l|}{\begin{tabular}[c]{@{}l@{}}Initial\\ state\end{tabular}} & \multicolumn{1}{l|}{\begin{tabular}[c]{@{}l@{}}$\left|DD\right>$ \\ (\%)\end{tabular}} & \multicolumn{1}{l|}{\begin{tabular}[c]{@{}l@{}}$\left|DS\right>$ \\ (\%)\end{tabular}} & \multicolumn{1}{l|}{\begin{tabular}[c]{@{}l@{}}$\left|SD\right>$ \\ (\%)\end{tabular}} & \multicolumn{1}{l|}{\begin{tabular}[c]{@{}l@{}}$\left|SS\right>$ \\ (\%)\end{tabular}} \\ \hline
\multirow{2}{*}{0}                                              & \multicolumn{1}{l|}{$\left|SD\right>$}                                       & \multicolumn{1}{l|}{$6 \pm 3$}            & \multicolumn{1}{l|}{$94 \pm 3$}           & \multicolumn{1}{l|}{$0 \pm 1$}         & \multicolumn{1}{l|}{$0 \pm 1$}         \\ \cline{3-6} 
                                                                & \multicolumn{1}{l|}{$\left|DS\right>$}                                       & \multicolumn{1}{l|}{$10 \pm 5$}        & \multicolumn{1}{l|}{$0 \pm 2$}         & \multicolumn{1}{l|}{$90 \pm 5$}        & \multicolumn{1}{l|}{$0 \pm 2$}         \\ \hline
\multirow{2}{*}{1}                                              & \multicolumn{1}{l|}{$\left|SD\right>$}                                       & \multicolumn{1}{l|}{$2 \pm 2$}         & \multicolumn{1}{l|}{$0 \pm 2$}         & \multicolumn{1}{l|}{$98 \pm 2$}        & \multicolumn{1}{l|}{$0 \pm 2$}         \\ \cline{3-6} 
                                                                & \multicolumn{1}{l|}{$\left|DS\right>$}                                       & \multicolumn{1}{l|}{$5 \pm 3$}         & \multicolumn{1}{l|}{$95 \pm 3$}        & \multicolumn{1}{l|}{$0 \pm 2$}         & \multicolumn{1}{l|}{$0 \pm 2$}         \\ \hline
\multirow{2}{*}{2}                                              & \multicolumn{1}{l|}{$\left|SD\right>$}                                       & \multicolumn{1}{l|}{$9 \pm 4$}         & \multicolumn{1}{l|}{$91 \pm 4$}        & \multicolumn{1}{l|}{$0 \pm 2$}         & \multicolumn{1}{l|}{$0 \pm 2$}         \\ \cline{3-6} 
                                                                & \multicolumn{1}{l|}{$\left|DS\right>$}                                       & \multicolumn{1}{l|}{$10 \pm 4$}        & \multicolumn{1}{l|}{$0 \pm 2$}         & \multicolumn{1}{l|}{$90 \pm 4$}        & \multicolumn{1}{l|}{$0 \pm 2$}         \\ \hline
\multirow{2}{*}{3}                                              & \multicolumn{1}{l|}{$\left|SD\right>$}                                       & \multicolumn{1}{l|}{$12 \pm 4$}        & \multicolumn{1}{l|}{$0 \pm 1$}         & \multicolumn{1}{l|}{$88 \pm 4$}        & \multicolumn{1}{l|}{$0 \pm 1$}         \\ \cline{3-6} 
                                                                & \multicolumn{1}{l|}{$\left|DS\right>$}                                       & \multicolumn{1}{l|}{$8 \pm 4$}         & \multicolumn{1}{l|}{$92 \pm 4$}        & \multicolumn{1}{l|}{$0 \pm 1$}         & \multicolumn{1}{l|}{$0 \pm 1$}         \\ \hline
\multirow{2}{*}{4}                                              & \multicolumn{1}{l|}{$\left|SD\right>$}                                       & \multicolumn{1}{l|}{$6 \pm 3$}         & \multicolumn{1}{l|}{$94 \pm 3$}        & \multicolumn{1}{l|}{$0 \pm 1$}         & \multicolumn{1}{l|}{$0 \pm 1$}         \\ \cline{3-6} 
                                                                & \multicolumn{1}{l|}{$\left|DS\right>$}                                       & \multicolumn{1}{l|}{$6 \pm 3$}         & \multicolumn{1}{l|}{$0 \pm 1$}         & \multicolumn{1}{l|}{$94 \pm 3$}        & \multicolumn{1}{l|}{$0 \pm 1$}         \\ \hline
\multirow{2}{*}{5}                                              & \multicolumn{1}{l|}{$\left|SD\right>$}                                       & \multicolumn{1}{l|}{$12 \pm 4$}        & \multicolumn{1}{l|}{$0 \pm 1$}         & \multicolumn{1}{l|}{$88 \pm 4$}        & \multicolumn{1}{l|}{$0 \pm 1$}         \\ \cline{3-6} 
                                                                & \multicolumn{1}{l|}{$\left|DS\right>$}                                       & \multicolumn{1}{l|}{$3 \pm 3$}         & \multicolumn{1}{l|}{$97 \pm 3$}        & \multicolumn{1}{l|}{$0 \pm 2$}         & \multicolumn{1}{l|}{$0 \pm 2$}         \\ \hline
\end{tabular}
\end{table}

%% file: outlook.tex
In this work we discussed a framework to generate the potentials required for rotations of single and multi-species ion crystals. A spherical harmonic expansion of the applied potentials serves as a natural basis to control trapping potentials for rotations. We described how these potentials can be determined and calibrated, and how they can be utilized to generate voltage sequences that rotate ion crystals. Furthermore, we demonstrated how machine learning methods can be used to overcome limitations in the achievable heating rate per rotation. These limitations are solely due to discrepancies between the apparatus and the physical model. We envision that this machine learning based approach in the field of applied optimal control theory may also find applications in the field of high-fidelity quantum control such as quantum simulations and quantum computation, as well as for quantum sensor applications to improve the achievable sensitivity.

The exchange of quantum information between qubits can also be realized by a SWAP gate acting on the qubit information, rather than exchanging the qubits themselves. Such a SWAP gate would be based on three CNOT gates. Given a fidelity of $~99.6\%$ \cite{Erhard_arXiv_2019}, a SWAP gate infidelity of $1-(.996)^3=1.2\%$ can be inferred. Compared to the infidelity achieved in the presented experiment of $0.2(2)\%$, we see a six-fold improvement compared to state-of-the-art quantum gates. This improvement demonstrates the benefits of ion transport based quantum information processing.

Concerning the machine learning based approach of optimal control, improvements can be made to the optimization algorithm to reduce both the final heating rate and the experimental time required to find it. For example, a more flexible searching algorithm can be designed that updates the search bounds based on previous measurements. For example, the DIRECT algorithm \cite{Finkel_OptOnl_2004} is designed to strike a compromise between exploring a broad, bounded parameter space, and focusing on converging on likely parameter candidates. Parameter regimes where voltage sequences might cause ion expulsion can be adaptively avoided. Also, periodic checks of rotation success can be performed to avoid an optimization of a non-rotating sequence.

Upon analysis of mode frequencies of our optimized rotation sequence, we note that the $X_{str}$ and $Z_{str}$ modes cross several times. The exchange of phonons between these modes is a dominant source of heating, and is reduced by cooling the $X_{str}$ mode before a rotation sequence. Ideally, one would want to cool all modes close to their ground states. Here, electromagnetically induced transparency (EIT) cooling \cite{Morigi_PhysRevLett_2000} and polarization gradient cooling \cite{Ejtemaee_PhysRevLett_2017} would be suitable approaches.

We intend to implement the suggested improvements in a next generation segmented slotted planar trap. The new trap has a 35\% lower electrode size to electrode-ion separation ratio, meaning a finer control over local trapping potentials. The filter design is upgraded to lower voltage crosstalk among the electrodes.
We expect these improvements to reduce heating of mixed species crystal rotations such that they remain in a regime of about $\bar{n}<10$ quanta, where sympathetically recooling the crystal to the ground state becomes viable.

This work presents the fundamentals and proof-of-principles of quantum computational protocols, in which ion transport reduces overhead and infidelity for quantum information processing. While the methods described in this work discuss the physical operation of swapping ions trapped in a planar trap, the methods can easily be generalized to other architectures and other physical operations. Characterization and expansion of potentials in terms of spherical harmonics is universally applicable to any Paul trap structure. The machine learning technique used for optimizing voltage sequences is also not restricted to ion crystal rotations; operations such as shutting, splitting, and merging ion crystals can benefit from machine learning algorithms for optimization.

%% file: acknowledgements.tex
We thank U. Poschinger for valuable discussions. This research was funded by the Office of the Director of National Intelligence (ODNI), Intelligence Advanced Research Projects Activity (IARPA), through the Army Research Office grant W911NF-16-1-0070. All statements of fact, opinions or conclusions contained herein are those of the authors and should not be construed as representing the official views or policies of IARPA, the ODNI, or the U.S. Government. We gratefully acknowledge support by the Austrian Science Fund (FWF), through the SFB BeyondC (FWF Project No. F71).

%% file: appendix_equil_pos.tex
In this appendix we show how to analytically find the equilibrium position for the case most relevant to this work, i.e. two ions. In order to simulate the dynamics of an ion crystal we find its equilibrium configuration under an external applied potential. The equilibrium positions $(x^{i}_1, x^{i}_2, x^{i}_3)$ of ions $i = 1,  \ldots, N$ in a crystal can be determined by minimizing their potential energy, which can be achieved numerically. However, it is computationally challenging to numerically design optimal trajectories under some parameterized external potential. It is therefore beneficial to use an analytical method, outlined below.

Let us first group the positions of the ions in the vector:
\begin{equation}
  \vect{x} = \left( \begin{matrix}
      x_1^{(1)} \\
      x_2^{(1)} \\ 
      x_3^{(1)} \\
      x_1^{(2)} \\
      x_2^{(2)} \\
      x_3^{(2)} 
    \end{matrix} \right).
\end{equation}
The potential energy of the crystal under an external harmonic potential is given by 
\begin{equation}
E_{\text{pot}} = \frac{1}{2} \vect{x}^T V \vect{x} + \vect{f}^T \cdot \vect{x} + \frac{e^2}{4 \pi \epsilon_0} r^{-1},
\end{equation}
where $V$ is the matrix describing the curvature of the external potential, $\vect{f}$ the vector describing the external electric field potential and $r$ is the distance between the ions. This distance can be written as:
\begin{equation}
  \label{eq:distance-two-ions}
  r^2 = \vect{x}^T C \vect{x},
\end{equation}
where:
\begin{equation}
  \label{eq:coulomb-matrix}
  C = \left(
  \begin{matrix}
    1 & 0 & 0 & -1 & 0 & 0\\
    0 & 1 & 0 & 0 & -1 & 0\\
    0 & 0 & 1 & 0 & 0 & -1\\
    -1 & 0 & 0 & 1 & 0 & 0\\
    0 & -1 & 0 & 0 & 1 & 0\\
    0 & 0 & -1 & 0 & 0 & 1
  \end{matrix} \right).
\end{equation}

Minimizing the potential energy is equivalent to solving the equation:
\begin{equation}
  \label{eq:zero-jacobian}
  \nabla E_{\text{pot}} = 0.
\end{equation}
Let us note that:
\begin{equation}
  \nabla(r^{-1}) = - r^{-3} C \vect{x},
\end{equation}
so (\ref{eq:zero-jacobian}) can be rewritten as:
\begin{equation}
  \label{eq:generalized-eigenvalue-problem}
  \left( V - \frac{e^2}{4 \pi \epsilon_0} r^{-3} C \right) \vect{x} = \vect{f}.
\end{equation}
This is a generalized eigenvalue problem with an inhomogeneous term $\vect{f}$, whose solutions $\vect{x}$ are the equilibrium positions.

To solve this generalized eigenvalue problem, we can find a change of coordinates $M$ that simultaneously diagonalizes $V$ and $C$, following the procedure explained in \cite{Bunse_gerstner_1984}. Multiplying (\ref{eq:generalized-eigenvalue-problem}) to the left by $M^T$ we obtain:
\begin{equation}
  \label{eq:transformed-eigenvalue-problem}
  \left( V' - \lambda C' \right) \vect{x'} = \vect{f'},
\end{equation}
where:
\begin{align}
  V' &= M^T V M \nonumber\\
  &= \left( \begin{matrix}
    v'_1 & 0 & 0\\
    0 & \ddots & 0\\
    0 & 0 & v'_6
  \end{matrix} \right),\\
  C' &= M^T C M \nonumber\\
  &= 2 \left(
  \begin{matrix}
    \mathbbm{1}_{3 \times 3} & 0_{3 \times 3} \\
    0_{3 \times 3} & 0_{3 \times 3}
  \end{matrix} \right) \\
  \label{eq:new-x}
  \vect{x'} &= M^{-1} \vect{x},\\
  \vect{f'} &= M^{T} \vect{f},
\end{align}
and the eigenvalue $\lambda$ is a rescaled distance:
\begin{equation}
  \label{eq:lambda}
  \lambda = \frac{e^2}{4 \pi \epsilon_0} r^{-3},
\end{equation}
subject to the constraint (\ref{eq:distance-two-ions}), which in the new coordinates reads:
\begin{equation}
\label{eq:rescaled-constraint}
\left( \frac{e^2}{4 \pi \epsilon_0} \right)^{2/3} \lambda^{-2/3} = 2 \sum_{i = 1}^3 (x'_i)^2.
\end{equation}

Since the new matrices $V'$ and $C'$ are diagonal, the left-hand side of (\ref{eq:transformed-eigenvalue-problem}) can be inverted efficiently. This yields the equilibrium positions in the new coordinates, which as a function of $\lambda$ are:
\begin{align}
  x'_i &= \begin{cases}
    f'_i / (v'_i - 2 \lambda) & \text{for } i = 1, 2, 3,\\
    f'_i / v'_i & \text{for } i = 4, 5, 6.\\
\end{cases}
\end{align}
The eigenvalue $\lambda$ can be found by replacing $x'_i$ into (\ref{eq:rescaled-constraint}), which yields a single scalar equation for $\lambda$:
\begin{align}
  \left( \frac{e^2}{4 \pi \epsilon_0} \right)^{2/3} \lambda^{-2/3} = \sum_{i = 1}^3 \frac{2 {f'}^2_i}{(v'_i - 2 \lambda)^2}.
\end{align}
This equation will in general have multiple solutions for $\lambda$, which correspond to all the equilibrium positions of the ions, stable or unstable. The stable equilibrium position can be found by calculating $E_{\text{pot}}$ for each $\lambda$ and finding the minimum. Finally, reversing (\ref{eq:new-x}) we can recover the equilibrium position $\vect{x}$ in the original coordinates.

%% file: appendix_voltage_solutions.tex
As described in Section \ref{sec:Rotation_potentials}, we parametrize our rotations in terms of time dependent spherical harmonic potentials. In our experiment, these potentials must be mapped to voltages applied to the trap electrodes. Here we describe how the voltage solutions for given potentials are determined.

Using boundary-element methods \cite{Singer_RevModPhys_2009}, we calculate the electrostatic potential $\Phi_{k,i}(x_i,y_i,z_i)$ generated by applying 1 volt to electrode $k$ at $n_i$ discrete points $\lbrace x_i,y_i,z_i\rbrace$ centered around the trapping region. The range of this grid of points is chosen such that higher-than-second order gradient terms are negligible. In our experiment we choose a range that is approximately 20\% of the electrode-ion separation, ie. $\approx$~20~ $\mu$m. The potential generated by the applied static voltages at the trap is given by $\sum_k \alpha_k \Phi_{k,i}(x_i,y_i,z_i)$, where $\alpha_k$ is the voltage applied to electrode $k$. The potential generated around the trapping region is to be expanded into the $n_m=8$ spherical harmonic terms of equation \ref{eq:spherical_harmonics} for all $K$ electrodes. 
To determine these electrode multipole expansions, first the potential of the $n_m$ spherical harmonic terms is calculated for a grid of points $\lbrace x_i,y_i,z_i\rbrace$, and put in an $n_m \times n_i$ matrix $\bm{\Psi}$ with terms $\Psi_{m,i}$.
The potential generated by each electrode is expanded into spherical harmonic terms by calculating the least-squares solution for $\textbf{M}$ in $\textbf{M}\bm{\Phi}=\bm{\Psi}$. Here, $\textbf{M}$ is a $(n_m\times K)$ matrix that contains the $n_m$ multipole terms of all $K$ electrodes, and $\bm{\Phi}$ is a $(K \times n_i)$ matrix containing the elements $\Phi_{k,i}$.
The set of voltages required to produce each harmonic potential is given by the singular-value decomposed pseudo-inverse $\textbf{M}^{-1}$, or in other words, the least-squares solution of the linear equation $\textbf{M}\textbf{x}_{K\times n_m}=\mathbbm{1}_{n_m}$.

We aim to generate potentials parametrized by the first 8 spherical harmonic coefficients, corresponding to $m_{l,n}$ from Section \ref{sec:exp_overview}, grouped in the matrix $\textbf{m}$.
In our setup, we control the trapping potentials by applying the voltages given by $\bm{\alpha}=\textbf{M}^{-1}\textbf{m}$.

%% file: appendix_JC.tex
\newcommand{\rzero}{\left|n\right\rangle\left|00\right\rangle}
\newcommand{\rone}{\left|n-1\right\rangle\left|10\right\rangle}
\newcommand{\reno}{\left|n-1\right\rangle\left|01\right\rangle}
\newcommand{\rtwo}{\left|n-2\right\rangle\left|11\right\rangle}
\newcommand{\bzero}{\left|n\right\rangle\left|00\right\rangle}
\newcommand{\bone}{\left|n+1\right\rangle\left|10\right\rangle}
\newcommand{\beno}{\left|n+1\right\rangle\left|01\right\rangle}
\newcommand{\btwo}{\left|n+2\right\rangle\left|11\right\rangle}
In this appendix we derive an analytic expression that describes excitations of motional sidebands of a two-ion crystal. The time-dependent excitations provide an accurate way to calculate mean phonon numbers of individual modes, since their evolutions are sensitive to the mean phonon number.

Measuring mode occupation numbers $\bar{n}$ for low phonon numbers ($\bar{n}\lesssim10$) is done by addressing motional sidebands at frequency $\omega=\omega_0 \pm \nu$, with $\omega_0$ the frequency of a $\left|S\right\rangle \leftrightarrow\left|D\right\rangle$ carrier transition, and $\nu$ the motional mode frequency.
The Jaynes-Cummings Hamiltonian
\begin{equation}
\hat{H}_{I_{\pm}} = \frac{1}{2} \hbar \Omega_{n\mp 1,n} \sum_i \left(\hat{n}\sigma_i^{\pm} + \hat{n}^{\dagger} \sigma_i^{\mp}\right),
\end{equation}
describes the interaction of this field with a red (blue) motional sideband, where the motional state $\left|n\right\rangle$ is occupied by $n$ phonons. Here, $\Omega_{n-1,n} = \eta\sqrt{n}\Omega$ and $\Omega_{n+1,n} = \eta\sqrt{n+1}\Omega$  are the blue and red sideband coupling strengths, $\sigma_i^+ = \left|1\right> \left<0\right|$ and $\sigma_i^- = \left|0\right> \left<1\right|$ the atomic transition operators acting on ion $i$, $\hat{a}$ and $\hat{a}^{\dagger}$ annihilation and creation operators. $\eta$ is the mode-frequency dependent Lamb-Dicke parameter, in our case $\eta=0.061$ and $0.047$ for the axial common and stretch modes. $H_{I_+}$ and $H_{I_-}$ represent interaction Hamiltonians on the blue and red sidebands, respectively.

The generalized ion state $\left|\phi(t)\right\rangle$  for two ions is given by
\begin{equation}
\begin{aligned}
\left|\phi(t)\right\rangle = \sum_{n=0}^{\infty} 
&a_{n}^{(0)} \left|n\right\rangle\left|00\right\rangle +
a_{n}^{(1)} \left|n\right\rangle\left|10\right\rangle +\\
&a_{n}^{(2)} \left|n\right\rangle\left|01\right\rangle +
a_{n}^{(3)} \left|n\right\rangle\left|11\right\rangle.
\end{aligned}
\end{equation}
with time-dependent coefficients $a_{n}^{(j)}$ obeying $\sum_{n}\sum_{j}|a_{n}^{(j)}|^2=1$. Given a symmetric starting state ($\left.a_n^{(1)}\right|_{t=0}=\left.a_n^{(2)}\right|_{t=0}$) and solely global operations, we can assume that $a_{n}^{(1)}=a_{n}^{(2)}$ for all $t$.
Applying a beam on resonance with a red motional sideband transition to a pair of ions in the optical ground state with phonon number $n$, $\left|\phi(0)\right\rangle=\left|n\right\rangle\left|00\right\rangle$, the system will remain in the subspace
\begin{equation}
\begin{aligned}
\left|\phi(t)\right\rangle =& 
\alpha_{00}\rzero+
\alpha_{01}\rone+\\
&\alpha_{10}\reno+
\alpha_{11}\rtwo.
\end{aligned}
\end{equation}
The time-dependent Schr\"odinger equation $i\hbar\frac{\partial}{\partial t}\left|\phi(t)\right\rangle=\hat{H}_{I_{-}}\left|\phi(t)\right\rangle$ is then:
\begin{equation}
\begin{aligned}
i\hbar(\dot{\alpha}_{00}\rzero+
\dot{\alpha}_{01}\rone+\\
\dot{\alpha}_{10}\reno+
\dot{\alpha}_{11}\rtwo) =\\
\frac{1}{2}\hbar\Omega_{n-1,n}\left[\alpha_{00}\left(\sqrt{n}\rone+\sqrt{n}\reno\right)\right.+\\
2\alpha_{01}\left(\sqrt{n-1}\rtwo+\sqrt{n}\rzero\right)+\\
\left.\alpha_{11}\left(\sqrt{n-1}\reno+\sqrt{n-1}\rone\right)\right]
\end{aligned}
\end{equation}
Combining identical states (dropping the index on $\Omega$ for convenience) gives a set of coupled first order differential equations:
\begin{equation}
\begin{aligned}
i\dot{\alpha}_{00}&=\Omega\sqrt{n}\alpha_{01}\\
2i\dot{\alpha}_{01}&=\Omega\left(\sqrt{n}\alpha_{00}+\sqrt{n-1}\alpha_{11}\right)\\
i\dot{\alpha}_{11}&=\Omega\sqrt{n-1}\alpha_{01}
\end{aligned}
\end{equation}
For initial conditions $\alpha_{00}=1$, $\alpha_{01}=\alpha_{11}=0$, the solutions are given by:
\begin{equation}
\begin{aligned}
\alpha_{00}(t)&=\frac{n-1+n\cos\left(\sqrt{4n-2}\Omega t\right)}{2n-1}\\
\alpha_{01}(t)&=-i\frac{\sqrt{2n^2-n}\sin\left(\sqrt{4n-2}\Omega t\right)}{\sqrt{2}(2n-1)}\\
\alpha_{11}(t)&=-\frac{\sqrt{n^2-n}\sin\left(\sqrt{4n-2}\Omega t\right)}{2n-1}
\end{aligned}
\end{equation}
A similar method is used to find the system equations for applying a beam on resonance with the blue motional sideband.

The excitation for two ions with $n$ phonons is given by $E_n(t)=\left|\alpha_{11}(t)\right|^2+0.5\left|\alpha_{01}(t)\right|^2$.
A thermally distributed mode has a phonon number $n$ occupation probability given by
\begin{equation}
P_{\bar{n}}(n) = \frac{1}{\bar{n}+1}\left(\frac{\bar{n}}{\bar{n}+1}\right)^n,
\end{equation}
for mean phonon number $\bar{n}$. The excitation for a thermally distributed mode is then
\begin{equation}
E_{\bar{n}}(t)=\sum_{n=0}^{\infty}P_{\bar{n}}(n)E_n(t)
\end{equation}
$\Omega_{n\pm1,1}$ is calculated with equation (\ref{eq:coupling_constants}), with $\Omega_{0,0}$ determined from a ground state-cooled carrier Rabi oscillation. $E_{\bar{n}}(t)$ is fit to experimental data of pulses on resonance with red and blue motional sidebands to determine the mean phonon number.

%% file: main.bbl
\begin{thebibliography}{52}%
\makeatletter
\providecommand \@ifxundefined [1]{%
 \@ifx{#1\undefined}
}%
\providecommand \@ifnum [1]{%
 \ifnum #1\expandafter \@firstoftwo
 \else \expandafter \@secondoftwo
 \fi
}%
\providecommand \@ifx [1]{%
 \ifx #1\expandafter \@firstoftwo
 \else \expandafter \@secondoftwo
 \fi
}%
\providecommand \natexlab [1]{#1}%
\providecommand \enquote  [1]{``#1''}%
\providecommand \bibnamefont  [1]{#1}%
\providecommand \bibfnamefont [1]{#1}%
\providecommand \citenamefont [1]{#1}%
\providecommand \href@noop [0]{\@secondoftwo}%
\providecommand \href [0]{\begingroup \@sanitize@url \@href}%
\providecommand \@href[1]{\@@startlink{#1}\@@href}%
\providecommand \@@href[1]{\endgroup#1\@@endlink}%
\providecommand \@sanitize@url [0]{\catcode `\\12\catcode `\$12\catcode
  `\&12\catcode `\#12\catcode `\^12\catcode `\_12\catcode `\%12\relax}%
\providecommand \@@startlink[1]{}%
\providecommand \@@endlink[0]{}%
\providecommand \url  [0]{\begingroup\@sanitize@url \@url }%
\providecommand \@url [1]{\endgroup\@href {#1}{\urlprefix }}%
\providecommand \urlprefix  [0]{URL }%
\providecommand \Eprint [0]{\href }%
\providecommand \doibase [0]{http://dx.doi.org/}%
\providecommand \selectlanguage [0]{\@gobble}%
\providecommand \bibinfo  [0]{\@secondoftwo}%
\providecommand \bibfield  [0]{\@secondoftwo}%
\providecommand \translation [1]{[#1]}%
\providecommand \BibitemOpen [0]{}%
\providecommand \bibitemStop [0]{}%
\providecommand \bibitemNoStop [0]{.\EOS\space}%
\providecommand \EOS [0]{\spacefactor3000\relax}%
\providecommand \BibitemShut  [1]{\csname bibitem#1\endcsname}%
\let\auto@bib@innerbib\@empty
\bibitem [{\citenamefont {Ladd}\ \emph {et~al.}(2010)\citenamefont {Ladd},
  \citenamefont {Jelezko}, \citenamefont {Laflamme}, \citenamefont {Nakamura},
  \citenamefont {Monroe},\ and\ \citenamefont {O'Brien}}]{Ladd_Nature_2010}%
  \BibitemOpen
  \bibfield  {author} {\bibinfo {author} {\bibfnamefont {T.~D.}\ \bibnamefont
  {Ladd}}, \bibinfo {author} {\bibfnamefont {F.}~\bibnamefont {Jelezko}},
  \bibinfo {author} {\bibfnamefont {R.}~\bibnamefont {Laflamme}}, \bibinfo
  {author} {\bibfnamefont {Y.}~\bibnamefont {Nakamura}}, \bibinfo {author}
  {\bibfnamefont {C.}~\bibnamefont {Monroe}}, \ and\ \bibinfo {author}
  {\bibfnamefont {J.~L.}\ \bibnamefont {O'Brien}},\ }\href
  {http://dx.doi.org/10.1038/nature08812} {\bibfield  {journal} {\bibinfo
  {journal} {Nature}\ }\textbf {\bibinfo {volume} {464}},\ \bibinfo {pages} {45
  EP } (\bibinfo {year} {2010})}\BibitemShut {NoStop}%
\bibitem [{\citenamefont {Wang}\ \emph {et~al.}(2017)\citenamefont {Wang},
  \citenamefont {Um}, \citenamefont {Zhang}, \citenamefont {An}, \citenamefont
  {Lyu}, \citenamefont {Zhang}, \citenamefont {Duan}, \citenamefont {Yum},\
  and\ \citenamefont {Kim}}]{Wang_NaturePhotonics_2017}%
  \BibitemOpen
  \bibfield  {author} {\bibinfo {author} {\bibfnamefont {Y.}~\bibnamefont
  {Wang}}, \bibinfo {author} {\bibfnamefont {M.}~\bibnamefont {Um}}, \bibinfo
  {author} {\bibfnamefont {J.}~\bibnamefont {Zhang}}, \bibinfo {author}
  {\bibfnamefont {S.}~\bibnamefont {An}}, \bibinfo {author} {\bibfnamefont
  {M.}~\bibnamefont {Lyu}}, \bibinfo {author} {\bibfnamefont {J.-N.}\
  \bibnamefont {Zhang}}, \bibinfo {author} {\bibfnamefont {L.-M.}\ \bibnamefont
  {Duan}}, \bibinfo {author} {\bibfnamefont {D.}~\bibnamefont {Yum}}, \ and\
  \bibinfo {author} {\bibfnamefont {K.}~\bibnamefont {Kim}},\ }\href {\doibase
  10.1038/s41566-017-0007-1} {\bibfield  {journal} {\bibinfo  {journal} {Nature
  Photonics}\ }\textbf {\bibinfo {volume} {11}},\ \bibinfo {pages} {646}
  (\bibinfo {year} {2017})}\BibitemShut {NoStop}%
\bibitem [{\citenamefont {Ballance}\ \emph {et~al.}(2016)\citenamefont
  {Ballance}, \citenamefont {Harty}, \citenamefont {Linke}, \citenamefont
  {Sepiol},\ and\ \citenamefont {Lucas}}]{Ballance_PhysRevLett_2016}%
  \BibitemOpen
  \bibfield  {author} {\bibinfo {author} {\bibfnamefont {C.}~\bibnamefont
  {Ballance}}, \bibinfo {author} {\bibfnamefont {T.}~\bibnamefont {Harty}},
  \bibinfo {author} {\bibfnamefont {N.}~\bibnamefont {Linke}}, \bibinfo
  {author} {\bibfnamefont {M.}~\bibnamefont {Sepiol}}, \ and\ \bibinfo {author}
  {\bibfnamefont {D.}~\bibnamefont {Lucas}},\ }\href@noop {} {\bibfield
  {journal} {\bibinfo  {journal} {Phys. Rev. Lett.}\ }\textbf {\bibinfo
  {volume} {117}},\ \bibinfo {pages} {060504} (\bibinfo {year}
  {2016})}\BibitemShut {NoStop}%
\bibitem [{\citenamefont {Gaebler}\ \emph {et~al.}(2016)\citenamefont
  {Gaebler}, \citenamefont {Tan}, \citenamefont {Lin}, \citenamefont {Wan},
  \citenamefont {Bowler}, \citenamefont {Keith}, \citenamefont {Glancy},
  \citenamefont {Coakley}, \citenamefont {Knill}, \citenamefont {Leibfried}
  \emph {et~al.}}]{Gaebler_PhysRevLett_2016}%
  \BibitemOpen
  \bibfield  {author} {\bibinfo {author} {\bibfnamefont {J.~P.}\ \bibnamefont
  {Gaebler}}, \bibinfo {author} {\bibfnamefont {T.~R.}\ \bibnamefont {Tan}},
  \bibinfo {author} {\bibfnamefont {Y.}~\bibnamefont {Lin}}, \bibinfo {author}
  {\bibfnamefont {Y.}~\bibnamefont {Wan}}, \bibinfo {author} {\bibfnamefont
  {R.}~\bibnamefont {Bowler}}, \bibinfo {author} {\bibfnamefont {A.~C.}\
  \bibnamefont {Keith}}, \bibinfo {author} {\bibfnamefont {S.}~\bibnamefont
  {Glancy}}, \bibinfo {author} {\bibfnamefont {K.}~\bibnamefont {Coakley}},
  \bibinfo {author} {\bibfnamefont {E.}~\bibnamefont {Knill}}, \bibinfo
  {author} {\bibfnamefont {D.}~\bibnamefont {Leibfried}},  \emph {et~al.},\
  }\href@noop {} {\bibfield  {journal} {\bibinfo  {journal} {Phys. Rev. Lett.}\
  }\textbf {\bibinfo {volume} {117}},\ \bibinfo {pages} {060505} (\bibinfo
  {year} {2016})}\BibitemShut {NoStop}%
\bibitem [{\citenamefont {Myerson}\ \emph {et~al.}(2008)\citenamefont
  {Myerson}, \citenamefont {Szwer}, \citenamefont {Webster}, \citenamefont
  {Allcock}, \citenamefont {Curtis}, \citenamefont {Imreh}, \citenamefont
  {Sherman}, \citenamefont {Stacey}, \citenamefont {Steane},\ and\
  \citenamefont {Lucas}}]{Myerson_PhysRevLett_2008}%
  \BibitemOpen
  \bibfield  {author} {\bibinfo {author} {\bibfnamefont {A.}~\bibnamefont
  {Myerson}}, \bibinfo {author} {\bibfnamefont {D.}~\bibnamefont {Szwer}},
  \bibinfo {author} {\bibfnamefont {S.}~\bibnamefont {Webster}}, \bibinfo
  {author} {\bibfnamefont {D.}~\bibnamefont {Allcock}}, \bibinfo {author}
  {\bibfnamefont {M.}~\bibnamefont {Curtis}}, \bibinfo {author} {\bibfnamefont
  {G.}~\bibnamefont {Imreh}}, \bibinfo {author} {\bibfnamefont
  {J.}~\bibnamefont {Sherman}}, \bibinfo {author} {\bibfnamefont
  {D.}~\bibnamefont {Stacey}}, \bibinfo {author} {\bibfnamefont
  {A.}~\bibnamefont {Steane}}, \ and\ \bibinfo {author} {\bibfnamefont
  {D.}~\bibnamefont {Lucas}},\ }\href@noop {} {\bibfield  {journal} {\bibinfo
  {journal} {Phys. Rev. Lett.}\ }\textbf {\bibinfo {volume} {100}},\ \bibinfo
  {pages} {200502} (\bibinfo {year} {2008})}\BibitemShut {NoStop}%
\bibitem [{\citenamefont {Turchette}\ \emph {et~al.}(2000)\citenamefont
  {Turchette}, \citenamefont {Kielpinski}, \citenamefont {King}, \citenamefont
  {Leibfried}, \citenamefont {Meekhof}, \citenamefont {Myatt}, \citenamefont
  {Rowe}, \citenamefont {Sackett}, \citenamefont {Wood}, \citenamefont {Itano},
  \citenamefont {Monroe},\ and\ \citenamefont
  {Wineland}}]{Turchette_PhysRevA_2000}%
  \BibitemOpen
  \bibfield  {author} {\bibinfo {author} {\bibfnamefont {Q.~A.}\ \bibnamefont
  {Turchette}}, \bibinfo {author} {\bibnamefont {Kielpinski}}, \bibinfo
  {author} {\bibfnamefont {B.~E.}\ \bibnamefont {King}}, \bibinfo {author}
  {\bibfnamefont {D.}~\bibnamefont {Leibfried}}, \bibinfo {author}
  {\bibfnamefont {D.~M.}\ \bibnamefont {Meekhof}}, \bibinfo {author}
  {\bibfnamefont {C.~J.}\ \bibnamefont {Myatt}}, \bibinfo {author}
  {\bibfnamefont {M.~A.}\ \bibnamefont {Rowe}}, \bibinfo {author}
  {\bibfnamefont {C.~A.}\ \bibnamefont {Sackett}}, \bibinfo {author}
  {\bibfnamefont {C.~S.}\ \bibnamefont {Wood}}, \bibinfo {author}
  {\bibfnamefont {W.~M.}\ \bibnamefont {Itano}}, \bibinfo {author}
  {\bibfnamefont {C.}~\bibnamefont {Monroe}}, \ and\ \bibinfo {author}
  {\bibfnamefont {D.~J.}\ \bibnamefont {Wineland}},\ }\href {\doibase
  10.1103/PhysRevA.61.063418} {\bibfield  {journal} {\bibinfo  {journal} {Phys.
  Rev. A}\ }\textbf {\bibinfo {volume} {61}},\ \bibinfo {pages} {063418}
  (\bibinfo {year} {2000})}\BibitemShut {NoStop}%
\bibitem [{\citenamefont {Brown}\ \emph {et~al.}(2016)\citenamefont {Brown},
  \citenamefont {Kim},\ and\ \citenamefont {Monroe}}]{Brown_Nature_2016}%
  \BibitemOpen
  \bibfield  {author} {\bibinfo {author} {\bibfnamefont {K.~R.}\ \bibnamefont
  {Brown}}, \bibinfo {author} {\bibfnamefont {J.}~\bibnamefont {Kim}}, \ and\
  \bibinfo {author} {\bibfnamefont {C.}~\bibnamefont {Monroe}},\ }\href@noop {}
  {\bibfield  {journal} {\bibinfo  {journal} {npj Quantum Information}\
  }\textbf {\bibinfo {volume} {2}},\ \bibinfo {pages} {16034} (\bibinfo {year}
  {2016})}\BibitemShut {NoStop}%
\bibitem [{\citenamefont {Kielpinski}\ \emph {et~al.}(2002)\citenamefont
  {Kielpinski}, \citenamefont {Monroe},\ and\ \citenamefont
  {Wineland}}]{Kielpinski_Nature_2002}%
  \BibitemOpen
  \bibfield  {author} {\bibinfo {author} {\bibfnamefont {D.}~\bibnamefont
  {Kielpinski}}, \bibinfo {author} {\bibfnamefont {C.}~\bibnamefont {Monroe}},
  \ and\ \bibinfo {author} {\bibfnamefont {D.}~\bibnamefont {Wineland}},\
  }\href@noop {} {\bibfield  {journal} {\bibinfo  {journal} {Nature}\ }\textbf
  {\bibinfo {volume} {417}},\ \bibinfo {pages} {709} (\bibinfo {year}
  {2002})}\BibitemShut {NoStop}%
\bibitem [{\citenamefont {Hughes}\ \emph {et~al.}(1996)\citenamefont {Hughes},
  \citenamefont {James}, \citenamefont {Knill}, \citenamefont {Laflamme},\ and\
  \citenamefont {Petschek}}]{Hughes_PhysRevLett_1996}%
  \BibitemOpen
  \bibfield  {author} {\bibinfo {author} {\bibfnamefont {R.~J.}\ \bibnamefont
  {Hughes}}, \bibinfo {author} {\bibfnamefont {D.~F.~V.}\ \bibnamefont
  {James}}, \bibinfo {author} {\bibfnamefont {E.~H.}\ \bibnamefont {Knill}},
  \bibinfo {author} {\bibfnamefont {R.}~\bibnamefont {Laflamme}}, \ and\
  \bibinfo {author} {\bibfnamefont {A.~G.}\ \bibnamefont {Petschek}},\ }\href
  {\doibase 10.1103/PhysRevLett.77.3240} {\bibfield  {journal} {\bibinfo
  {journal} {Phys. Rev. Lett.}\ }\textbf {\bibinfo {volume} {77}},\ \bibinfo
  {pages} {3240} (\bibinfo {year} {1996})}\BibitemShut {NoStop}%
\bibitem [{\citenamefont {Hensinger}\ \emph {et~al.}(2006)\citenamefont
  {Hensinger}, \citenamefont {Olmschenk}, \citenamefont {Stick}, \citenamefont
  {Hucul}, \citenamefont {Yeo}, \citenamefont {Acton}, \citenamefont
  {Deslauriers}, \citenamefont {Monroe},\ and\ \citenamefont
  {Rabchuk}}]{Hensinger_APL_2006}%
  \BibitemOpen
  \bibfield  {author} {\bibinfo {author} {\bibfnamefont {W.~K.}\ \bibnamefont
  {Hensinger}}, \bibinfo {author} {\bibfnamefont {S.}~\bibnamefont
  {Olmschenk}}, \bibinfo {author} {\bibfnamefont {D.}~\bibnamefont {Stick}},
  \bibinfo {author} {\bibfnamefont {D.}~\bibnamefont {Hucul}}, \bibinfo
  {author} {\bibfnamefont {M.}~\bibnamefont {Yeo}}, \bibinfo {author}
  {\bibfnamefont {M.}~\bibnamefont {Acton}}, \bibinfo {author} {\bibfnamefont
  {L.}~\bibnamefont {Deslauriers}}, \bibinfo {author} {\bibfnamefont
  {C.}~\bibnamefont {Monroe}}, \ and\ \bibinfo {author} {\bibfnamefont
  {J.}~\bibnamefont {Rabchuk}},\ }\href {\doibase 10.1063/1.2164910} {\bibfield
   {journal} {\bibinfo  {journal} {Appl. Phys. Lett.}\ }\textbf {\bibinfo
  {volume} {88}},\ \bibinfo {pages} {034101} (\bibinfo {year}
  {2006})}\BibitemShut {NoStop}%
\bibitem [{\citenamefont {Hughes}\ \emph {et~al.}(2011)\citenamefont {Hughes},
  \citenamefont {Lekitsch}, \citenamefont {Broersma},\ and\ \citenamefont
  {Hensinger}}]{Hughes_ContPhys_2011}%
  \BibitemOpen
  \bibfield  {author} {\bibinfo {author} {\bibfnamefont {M.~D.}\ \bibnamefont
  {Hughes}}, \bibinfo {author} {\bibfnamefont {B.}~\bibnamefont {Lekitsch}},
  \bibinfo {author} {\bibfnamefont {J.~A.}\ \bibnamefont {Broersma}}, \ and\
  \bibinfo {author} {\bibfnamefont {W.~K.}\ \bibnamefont {Hensinger}},\
  }\href@noop {} {\bibfield  {journal} {\bibinfo  {journal} {Contemporary
  Physics}\ }\textbf {\bibinfo {volume} {52}},\ \bibinfo {pages} {505}
  (\bibinfo {year} {2011})}\BibitemShut {NoStop}%
\bibitem [{\citenamefont {Mack}(2008)}]{Mack_2008}%
  \BibitemOpen
  \bibfield  {author} {\bibinfo {author} {\bibfnamefont {C.}~\bibnamefont
  {Mack}},\ }\href@noop {} {\emph {\bibinfo {title} {Fundamental principles of
  optical lithography: the science of microfabrication}}}\ (\bibinfo
  {publisher} {John Wiley \& Sons},\ \bibinfo {year} {2008})\BibitemShut
  {NoStop}%
\bibitem [{\citenamefont {Seidelin}\ \emph {et~al.}(2006)\citenamefont
  {Seidelin}, \citenamefont {Chiaverini}, \citenamefont {Reichle},
  \citenamefont {Bollinger}, \citenamefont {Leibfried}, \citenamefont
  {Britton}, \citenamefont {Wesenberg}, \citenamefont {Blakestad},
  \citenamefont {Epstein}, \citenamefont {Hume}, \citenamefont {Itano},
  \citenamefont {Jost}, \citenamefont {Langer}, \citenamefont {Ozeri},
  \citenamefont {Shiga},\ and\ \citenamefont
  {Wineland}}]{Seidelin_PhysRevLett_2006}%
  \BibitemOpen
  \bibfield  {author} {\bibinfo {author} {\bibfnamefont {S.}~\bibnamefont
  {Seidelin}}, \bibinfo {author} {\bibfnamefont {J.}~\bibnamefont
  {Chiaverini}}, \bibinfo {author} {\bibfnamefont {R.}~\bibnamefont {Reichle}},
  \bibinfo {author} {\bibfnamefont {J.~J.}\ \bibnamefont {Bollinger}}, \bibinfo
  {author} {\bibfnamefont {D.}~\bibnamefont {Leibfried}}, \bibinfo {author}
  {\bibfnamefont {J.}~\bibnamefont {Britton}}, \bibinfo {author} {\bibfnamefont
  {J.~H.}\ \bibnamefont {Wesenberg}}, \bibinfo {author} {\bibfnamefont {R.~B.}\
  \bibnamefont {Blakestad}}, \bibinfo {author} {\bibfnamefont {R.~J.}\
  \bibnamefont {Epstein}}, \bibinfo {author} {\bibfnamefont {D.~B.}\
  \bibnamefont {Hume}}, \bibinfo {author} {\bibfnamefont {W.~M.}\ \bibnamefont
  {Itano}}, \bibinfo {author} {\bibfnamefont {J.~D.}\ \bibnamefont {Jost}},
  \bibinfo {author} {\bibfnamefont {C.}~\bibnamefont {Langer}}, \bibinfo
  {author} {\bibfnamefont {R.}~\bibnamefont {Ozeri}}, \bibinfo {author}
  {\bibfnamefont {N.}~\bibnamefont {Shiga}}, \ and\ \bibinfo {author}
  {\bibfnamefont {D.~J.}\ \bibnamefont {Wineland}},\ }\href {\doibase
  10.1103/PhysRevLett.96.253003} {\bibfield  {journal} {\bibinfo  {journal}
  {Phys. Rev. Lett.}\ }\textbf {\bibinfo {volume} {96}},\ \bibinfo {pages}
  {253003} (\bibinfo {year} {2006})}\BibitemShut {NoStop}%
\bibitem [{\citenamefont {Mehta}\ \emph {et~al.}(2014)\citenamefont {Mehta},
  \citenamefont {Eltony}, \citenamefont {Bruzewicz}, \citenamefont {Chuang},
  \citenamefont {Ram}, \citenamefont {Sage},\ and\ \citenamefont
  {Chiaverini}}]{Mehta_ApplPhysLett_2014}%
  \BibitemOpen
  \bibfield  {author} {\bibinfo {author} {\bibfnamefont {K.~K.}\ \bibnamefont
  {Mehta}}, \bibinfo {author} {\bibfnamefont {A.~M.}\ \bibnamefont {Eltony}},
  \bibinfo {author} {\bibfnamefont {C.~D.}\ \bibnamefont {Bruzewicz}}, \bibinfo
  {author} {\bibfnamefont {I.~L.}\ \bibnamefont {Chuang}}, \bibinfo {author}
  {\bibfnamefont {R.~J.}\ \bibnamefont {Ram}}, \bibinfo {author} {\bibfnamefont
  {J.~M.}\ \bibnamefont {Sage}}, \ and\ \bibinfo {author} {\bibfnamefont
  {J.}~\bibnamefont {Chiaverini}},\ }\href {\doibase 10.1063/1.4892061}
  {\bibfield  {journal} {\bibinfo  {journal} {Appl. Phys. Lett.}\ }\textbf
  {\bibinfo {volume} {105}},\ \bibinfo {pages} {044103} (\bibinfo {year}
  {2014})}\BibitemShut {NoStop}%
\bibitem [{\citenamefont {Schindler}\ \emph {et~al.}(2013)\citenamefont
  {Schindler}, \citenamefont {Nigg}, \citenamefont {Monz}, \citenamefont
  {Barreiro}, \citenamefont {Martinez}, \citenamefont {Wang}, \citenamefont
  {Quint}, \citenamefont {Brandl}, \citenamefont {Nebendahl}, \citenamefont
  {Roos}, \citenamefont {Chwalla}, \citenamefont {Hennrich},\ and\
  \citenamefont {Blatt}}]{Schindler_NJP_2013}%
  \BibitemOpen
  \bibfield  {author} {\bibinfo {author} {\bibfnamefont {P.}~\bibnamefont
  {Schindler}}, \bibinfo {author} {\bibfnamefont {D.}~\bibnamefont {Nigg}},
  \bibinfo {author} {\bibfnamefont {T.}~\bibnamefont {Monz}}, \bibinfo {author}
  {\bibfnamefont {J.~T.}\ \bibnamefont {Barreiro}}, \bibinfo {author}
  {\bibfnamefont {E.}~\bibnamefont {Martinez}}, \bibinfo {author}
  {\bibfnamefont {S.~X.}\ \bibnamefont {Wang}}, \bibinfo {author}
  {\bibfnamefont {S.}~\bibnamefont {Quint}}, \bibinfo {author} {\bibfnamefont
  {M.~F.}\ \bibnamefont {Brandl}}, \bibinfo {author} {\bibfnamefont
  {V.}~\bibnamefont {Nebendahl}}, \bibinfo {author} {\bibfnamefont {C.~F.}\
  \bibnamefont {Roos}}, \bibinfo {author} {\bibfnamefont {M.}~\bibnamefont
  {Chwalla}}, \bibinfo {author} {\bibfnamefont {M.}~\bibnamefont {Hennrich}}, \
  and\ \bibinfo {author} {\bibfnamefont {R.}~\bibnamefont {Blatt}},\
  }\href@noop {} {\bibfield  {journal} {\bibinfo  {journal} {New J. of Phys.}\
  }\textbf {\bibinfo {volume} {15}},\ \bibinfo {pages} {123012} (\bibinfo
  {year} {2013})}\BibitemShut {NoStop}%
\bibitem [{\citenamefont {Walther}\ \emph {et~al.}(2012)\citenamefont
  {Walther}, \citenamefont {Ziesel}, \citenamefont {Ruster}, \citenamefont
  {Dawkins}, \citenamefont {Ott}, \citenamefont {Hettrich}, \citenamefont
  {Singer}, \citenamefont {Schmidt-Kaler},\ and\ \citenamefont
  {Poschinger}}]{Walther_PhysRevLett_2012}%
  \BibitemOpen
  \bibfield  {author} {\bibinfo {author} {\bibfnamefont {A.}~\bibnamefont
  {Walther}}, \bibinfo {author} {\bibfnamefont {F.}~\bibnamefont {Ziesel}},
  \bibinfo {author} {\bibfnamefont {T.}~\bibnamefont {Ruster}}, \bibinfo
  {author} {\bibfnamefont {S.~T.}\ \bibnamefont {Dawkins}}, \bibinfo {author}
  {\bibfnamefont {K.}~\bibnamefont {Ott}}, \bibinfo {author} {\bibfnamefont
  {M.}~\bibnamefont {Hettrich}}, \bibinfo {author} {\bibfnamefont
  {K.}~\bibnamefont {Singer}}, \bibinfo {author} {\bibfnamefont
  {F.}~\bibnamefont {Schmidt-Kaler}}, \ and\ \bibinfo {author} {\bibfnamefont
  {U.}~\bibnamefont {Poschinger}},\ }\href {\doibase
  10.1103/PhysRevLett.109.080501} {\bibfield  {journal} {\bibinfo  {journal}
  {Phys. Rev. Lett.}\ }\textbf {\bibinfo {volume} {109}},\ \bibinfo {pages}
  {080501} (\bibinfo {year} {2012})}\BibitemShut {NoStop}%
\bibitem [{\citenamefont {Kaufmann}\ \emph {et~al.}(2014)\citenamefont
  {Kaufmann}, \citenamefont {Ruster}, \citenamefont {Schmiegelow},
  \citenamefont {Schmidt-Kaler},\ and\ \citenamefont
  {Poschinger}}]{Kaufmann_NJP_2014}%
  \BibitemOpen
  \bibfield  {author} {\bibinfo {author} {\bibfnamefont {H.}~\bibnamefont
  {Kaufmann}}, \bibinfo {author} {\bibfnamefont {T.}~\bibnamefont {Ruster}},
  \bibinfo {author} {\bibfnamefont {C.~T.}\ \bibnamefont {Schmiegelow}},
  \bibinfo {author} {\bibfnamefont {F.}~\bibnamefont {Schmidt-Kaler}}, \ and\
  \bibinfo {author} {\bibfnamefont {U.~G.}\ \bibnamefont {Poschinger}},\ }\href
  {http://stacks.iop.org/1367-2630/16/i=7/a=073012} {\bibfield  {journal}
  {\bibinfo  {journal} {New J. of Phys.}\ }\textbf {\bibinfo {volume} {16}},\
  \bibinfo {pages} {073012} (\bibinfo {year} {2014})}\BibitemShut {NoStop}%
\bibitem [{\citenamefont {Bowler}\ \emph {et~al.}(2012)\citenamefont {Bowler},
  \citenamefont {Gaebler}, \citenamefont {Lin}, \citenamefont {Tan},
  \citenamefont {Hanneke}, \citenamefont {Jost}, \citenamefont {Home},
  \citenamefont {Leibfried},\ and\ \citenamefont
  {Wineland}}]{Bowler_PhysRevLett_2012}%
  \BibitemOpen
  \bibfield  {author} {\bibinfo {author} {\bibfnamefont {R.}~\bibnamefont
  {Bowler}}, \bibinfo {author} {\bibfnamefont {J.}~\bibnamefont {Gaebler}},
  \bibinfo {author} {\bibfnamefont {Y.}~\bibnamefont {Lin}}, \bibinfo {author}
  {\bibfnamefont {T.~R.}\ \bibnamefont {Tan}}, \bibinfo {author} {\bibfnamefont
  {D.}~\bibnamefont {Hanneke}}, \bibinfo {author} {\bibfnamefont {J.~D.}\
  \bibnamefont {Jost}}, \bibinfo {author} {\bibfnamefont {J.}~\bibnamefont
  {Home}}, \bibinfo {author} {\bibfnamefont {D.}~\bibnamefont {Leibfried}}, \
  and\ \bibinfo {author} {\bibfnamefont {D.~J.}\ \bibnamefont {Wineland}},\
  }\href@noop {} {\bibfield  {journal} {\bibinfo  {journal} {Phys. Rev. Lett.}\
  }\textbf {\bibinfo {volume} {109}},\ \bibinfo {pages} {080502} (\bibinfo
  {year} {2012})}\BibitemShut {NoStop}%
\bibitem [{\citenamefont {Shu}\ \emph {et~al.}(2014)\citenamefont {Shu},
  \citenamefont {Vittorini}, \citenamefont {Buikema}, \citenamefont {Nichols},
  \citenamefont {Volin}, \citenamefont {Stick},\ and\ \citenamefont
  {Brown}}]{Shu_PhysRevA_2014}%
  \BibitemOpen
  \bibfield  {author} {\bibinfo {author} {\bibfnamefont {G.}~\bibnamefont
  {Shu}}, \bibinfo {author} {\bibfnamefont {G.}~\bibnamefont {Vittorini}},
  \bibinfo {author} {\bibfnamefont {A.}~\bibnamefont {Buikema}}, \bibinfo
  {author} {\bibfnamefont {C.~S.}\ \bibnamefont {Nichols}}, \bibinfo {author}
  {\bibfnamefont {C.}~\bibnamefont {Volin}}, \bibinfo {author} {\bibfnamefont
  {D.}~\bibnamefont {Stick}}, \ and\ \bibinfo {author} {\bibfnamefont {K.~R.}\
  \bibnamefont {Brown}},\ }\href {\doibase 10.1103/PhysRevA.89.062308}
  {\bibfield  {journal} {\bibinfo  {journal} {Phys. Rev. A}\ }\textbf {\bibinfo
  {volume} {89}},\ \bibinfo {pages} {062308} (\bibinfo {year}
  {2014})}\BibitemShut {NoStop}%
\bibitem [{\citenamefont {Blakestad}\ \emph {et~al.}(2009)\citenamefont
  {Blakestad}, \citenamefont {Ospelkaus}, \citenamefont {VanDevender},
  \citenamefont {Amini}, \citenamefont {Britton}, \citenamefont {Leibfried},\
  and\ \citenamefont {Wineland}}]{Blakestad_PhysRevLett_2009}%
  \BibitemOpen
  \bibfield  {author} {\bibinfo {author} {\bibfnamefont {R.~B.}\ \bibnamefont
  {Blakestad}}, \bibinfo {author} {\bibfnamefont {C.}~\bibnamefont
  {Ospelkaus}}, \bibinfo {author} {\bibfnamefont {A.~P.}\ \bibnamefont
  {VanDevender}}, \bibinfo {author} {\bibfnamefont {J.~M.}\ \bibnamefont
  {Amini}}, \bibinfo {author} {\bibfnamefont {J.}~\bibnamefont {Britton}},
  \bibinfo {author} {\bibfnamefont {D.}~\bibnamefont {Leibfried}}, \ and\
  \bibinfo {author} {\bibfnamefont {D.~J.}\ \bibnamefont {Wineland}},\ }\href
  {\doibase 10.1103/PhysRevLett.102.153002} {\bibfield  {journal} {\bibinfo
  {journal} {Phys. Rev. Lett.}\ }\textbf {\bibinfo {volume} {102}},\ \bibinfo
  {pages} {153002} (\bibinfo {year} {2009})}\BibitemShut {NoStop}%
\bibitem [{\citenamefont {Blakestad}\ \emph {et~al.}(2011)\citenamefont
  {Blakestad}, \citenamefont {Ospelkaus}, \citenamefont {VanDevender},
  \citenamefont {Wesenberg}, \citenamefont {Biercuk}, \citenamefont
  {Leibfried},\ and\ \citenamefont {Wineland}}]{Blakestad_PhysRevA_2011}%
  \BibitemOpen
  \bibfield  {author} {\bibinfo {author} {\bibfnamefont {R.~B.}\ \bibnamefont
  {Blakestad}}, \bibinfo {author} {\bibfnamefont {C.}~\bibnamefont
  {Ospelkaus}}, \bibinfo {author} {\bibfnamefont {A.~P.}\ \bibnamefont
  {VanDevender}}, \bibinfo {author} {\bibfnamefont {J.~H.}\ \bibnamefont
  {Wesenberg}}, \bibinfo {author} {\bibfnamefont {M.~J.}\ \bibnamefont
  {Biercuk}}, \bibinfo {author} {\bibfnamefont {D.}~\bibnamefont {Leibfried}},
  \ and\ \bibinfo {author} {\bibfnamefont {D.~J.}\ \bibnamefont {Wineland}},\
  }\href {\doibase 10.1103/PhysRevA.84.032314} {\bibfield  {journal} {\bibinfo
  {journal} {Phys. Rev. A}\ }\textbf {\bibinfo {volume} {84}},\ \bibinfo
  {pages} {032314} (\bibinfo {year} {2011})}\BibitemShut {NoStop}%
\bibitem [{\citenamefont {Wright}\ \emph {et~al.}(2013)\citenamefont {Wright},
  \citenamefont {Amini}, \citenamefont {Faircloth}, \citenamefont {Volin},
  \citenamefont {Doret}, \citenamefont {Hayden}, \citenamefont {Pai},
  \citenamefont {Landgren}, \citenamefont {Denison}, \citenamefont {Killian}
  \emph {et~al.}}]{Wright_NJP_2013}%
  \BibitemOpen
  \bibfield  {author} {\bibinfo {author} {\bibfnamefont {K.}~\bibnamefont
  {Wright}}, \bibinfo {author} {\bibfnamefont {J.~M.}\ \bibnamefont {Amini}},
  \bibinfo {author} {\bibfnamefont {D.~L.}\ \bibnamefont {Faircloth}}, \bibinfo
  {author} {\bibfnamefont {C.}~\bibnamefont {Volin}}, \bibinfo {author}
  {\bibfnamefont {S.~C.}\ \bibnamefont {Doret}}, \bibinfo {author}
  {\bibfnamefont {H.}~\bibnamefont {Hayden}}, \bibinfo {author} {\bibfnamefont
  {C.}~\bibnamefont {Pai}}, \bibinfo {author} {\bibfnamefont {D.~W.}\
  \bibnamefont {Landgren}}, \bibinfo {author} {\bibfnamefont {D.}~\bibnamefont
  {Denison}}, \bibinfo {author} {\bibfnamefont {T.}~\bibnamefont {Killian}},
  \emph {et~al.},\ }\href@noop {} {\bibfield  {journal} {\bibinfo  {journal}
  {New J. of Phys.}\ }\textbf {\bibinfo {volume} {15}},\ \bibinfo {pages}
  {033004} (\bibinfo {year} {2013})}\BibitemShut {NoStop}%
\bibitem [{\citenamefont {Kaufmann}\ \emph {et~al.}(2017)\citenamefont
  {Kaufmann}, \citenamefont {Ruster}, \citenamefont {Schmiegelow},
  \citenamefont {Luda}, \citenamefont {Kaushal}, \citenamefont {Schulz},
  \citenamefont {von Lindenfels}, \citenamefont {Schmidt-Kaler},\ and\
  \citenamefont {Poschinger}}]{Kaufmann_PRL_2017}%
  \BibitemOpen
  \bibfield  {author} {\bibinfo {author} {\bibfnamefont {H.}~\bibnamefont
  {Kaufmann}}, \bibinfo {author} {\bibfnamefont {T.}~\bibnamefont {Ruster}},
  \bibinfo {author} {\bibfnamefont {C.~T.}\ \bibnamefont {Schmiegelow}},
  \bibinfo {author} {\bibfnamefont {M.~A.}\ \bibnamefont {Luda}}, \bibinfo
  {author} {\bibfnamefont {V.}~\bibnamefont {Kaushal}}, \bibinfo {author}
  {\bibfnamefont {J.}~\bibnamefont {Schulz}}, \bibinfo {author} {\bibfnamefont
  {D.}~\bibnamefont {von Lindenfels}}, \bibinfo {author} {\bibfnamefont
  {F.}~\bibnamefont {Schmidt-Kaler}}, \ and\ \bibinfo {author} {\bibfnamefont
  {U.~G.}\ \bibnamefont {Poschinger}},\ }\href {\doibase
  10.1103/PhysRevA.95.052319} {\bibfield  {journal} {\bibinfo  {journal} {Phys.
  Rev. A}\ }\textbf {\bibinfo {volume} {95}},\ \bibinfo {pages} {052319}
  (\bibinfo {year} {2017})}\BibitemShut {NoStop}%
\bibitem [{\citenamefont {Splatt}\ \emph {et~al.}(2009)\citenamefont {Splatt},
  \citenamefont {Harlander}, \citenamefont {Brownnutt}, \citenamefont
  {Zähringer}, \citenamefont {Blatt},\ and\ \citenamefont
  {Hänsel}}]{Splatt_NJP_2009}%
  \BibitemOpen
  \bibfield  {author} {\bibinfo {author} {\bibfnamefont {F.}~\bibnamefont
  {Splatt}}, \bibinfo {author} {\bibfnamefont {M.}~\bibnamefont {Harlander}},
  \bibinfo {author} {\bibfnamefont {M.}~\bibnamefont {Brownnutt}}, \bibinfo
  {author} {\bibfnamefont {F.}~\bibnamefont {Zähringer}}, \bibinfo {author}
  {\bibfnamefont {R.}~\bibnamefont {Blatt}}, \ and\ \bibinfo {author}
  {\bibfnamefont {W.}~\bibnamefont {Hänsel}},\ }\href
  {http://stacks.iop.org/1367-2630/11/i=10/a=103008} {\bibfield  {journal}
  {\bibinfo  {journal} {New J. of Phys.}\ }\textbf {\bibinfo {volume} {11}},\
  \bibinfo {pages} {103008} (\bibinfo {year} {2009})}\BibitemShut {NoStop}%
\bibitem [{\citenamefont {Maunz}(2017)}]{Maunz_2017}%
  \BibitemOpen
  \bibfield  {author} {\bibinfo {author} {\bibfnamefont {P.}~\bibnamefont
  {Maunz}},\ }\href@noop {} {}\bibinfo {howpublished} {Private Communications}
  (\bibinfo {year} {2017})\BibitemShut {NoStop}%
\bibitem [{\citenamefont {Chiaverini}\ \emph {et~al.}(2005)\citenamefont
  {Chiaverini}, \citenamefont {Blakestad}, \citenamefont {Britton},
  \citenamefont {Jost}, \citenamefont {Langer}, \citenamefont {Leibfried},
  \citenamefont {Ozeri},\ and\ \citenamefont {Wineland}}]{Chiaverini_QIC_2005}%
  \BibitemOpen
  \bibfield  {author} {\bibinfo {author} {\bibfnamefont {J.}~\bibnamefont
  {Chiaverini}}, \bibinfo {author} {\bibfnamefont {R.}~\bibnamefont
  {Blakestad}}, \bibinfo {author} {\bibfnamefont {J.}~\bibnamefont {Britton}},
  \bibinfo {author} {\bibfnamefont {J.}~\bibnamefont {Jost}}, \bibinfo {author}
  {\bibfnamefont {C.}~\bibnamefont {Langer}}, \bibinfo {author} {\bibfnamefont
  {D.}~\bibnamefont {Leibfried}}, \bibinfo {author} {\bibfnamefont
  {R.}~\bibnamefont {Ozeri}}, \ and\ \bibinfo {author} {\bibfnamefont
  {D.}~\bibnamefont {Wineland}},\ }\href@noop {} {\bibfield  {journal}
  {\bibinfo  {journal} {Quantum Information \& Computation}\ }\textbf {\bibinfo
  {volume} {5}},\ \bibinfo {pages} {419} (\bibinfo {year} {2005})}\BibitemShut
  {NoStop}%
\bibitem [{\citenamefont {Bermudez}\ \emph {et~al.}(2017)\citenamefont
  {Bermudez}, \citenamefont {Xu}, \citenamefont {Nigmatullin}, \citenamefont
  {O'Gorman}, \citenamefont {Negnevitsky}, \citenamefont {Schindler},
  \citenamefont {Monz}, \citenamefont {Poschinger}, \citenamefont {Hempel},
  \citenamefont {Home}, \citenamefont {Schmidt-Kaler}, \citenamefont {Biercuk},
  \citenamefont {Blatt}, \citenamefont {Benjamin},\ and\ \citenamefont
  {M\"uller}}]{Bermudez_PhysRevX_2017}%
  \BibitemOpen
  \bibfield  {author} {\bibinfo {author} {\bibfnamefont {A.}~\bibnamefont
  {Bermudez}}, \bibinfo {author} {\bibfnamefont {X.}~\bibnamefont {Xu}},
  \bibinfo {author} {\bibfnamefont {R.}~\bibnamefont {Nigmatullin}}, \bibinfo
  {author} {\bibfnamefont {J.}~\bibnamefont {O'Gorman}}, \bibinfo {author}
  {\bibfnamefont {V.}~\bibnamefont {Negnevitsky}}, \bibinfo {author}
  {\bibfnamefont {P.}~\bibnamefont {Schindler}}, \bibinfo {author}
  {\bibfnamefont {T.}~\bibnamefont {Monz}}, \bibinfo {author} {\bibfnamefont
  {U.~G.}\ \bibnamefont {Poschinger}}, \bibinfo {author} {\bibfnamefont
  {C.}~\bibnamefont {Hempel}}, \bibinfo {author} {\bibfnamefont
  {J.}~\bibnamefont {Home}}, \bibinfo {author} {\bibfnamefont {F.}~\bibnamefont
  {Schmidt-Kaler}}, \bibinfo {author} {\bibfnamefont {M.}~\bibnamefont
  {Biercuk}}, \bibinfo {author} {\bibfnamefont {R.}~\bibnamefont {Blatt}},
  \bibinfo {author} {\bibfnamefont {S.}~\bibnamefont {Benjamin}}, \ and\
  \bibinfo {author} {\bibfnamefont {M.}~\bibnamefont {M\"uller}},\ }\href
  {\doibase 10.1103/PhysRevX.7.041061} {\bibfield  {journal} {\bibinfo
  {journal} {Phys. Rev. X}\ }\textbf {\bibinfo {volume} {7}},\ \bibinfo {pages}
  {041061} (\bibinfo {year} {2017})}\BibitemShut {NoStop}%
\bibitem [{\citenamefont {Brandl}\ \emph {et~al.}(2016)\citenamefont {Brandl},
  \citenamefont {van Mourik}, \citenamefont {Postler}, \citenamefont {Nolf},
  \citenamefont {Lakhmanskiy}, \citenamefont {Paiva}, \citenamefont {Möller},
  \citenamefont {Daniilidis}, \citenamefont {Häffner}, \citenamefont
  {Kaushal}, \citenamefont {Ruster}, \citenamefont {Warschburger},
  \citenamefont {Kaufmann}, \citenamefont {Poschinger}, \citenamefont
  {Schmidt-Kaler}, \citenamefont {Schindler}, \citenamefont {Monz},\ and\
  \citenamefont {Blatt}}]{Brandl_RSI_2016}%
  \BibitemOpen
  \bibfield  {author} {\bibinfo {author} {\bibfnamefont {M.~F.}\ \bibnamefont
  {Brandl}}, \bibinfo {author} {\bibfnamefont {M.~W.}\ \bibnamefont {van
  Mourik}}, \bibinfo {author} {\bibfnamefont {L.}~\bibnamefont {Postler}},
  \bibinfo {author} {\bibfnamefont {A.}~\bibnamefont {Nolf}}, \bibinfo {author}
  {\bibfnamefont {K.}~\bibnamefont {Lakhmanskiy}}, \bibinfo {author}
  {\bibfnamefont {R.~R.}\ \bibnamefont {Paiva}}, \bibinfo {author}
  {\bibfnamefont {S.}~\bibnamefont {Möller}}, \bibinfo {author} {\bibfnamefont
  {N.}~\bibnamefont {Daniilidis}}, \bibinfo {author} {\bibfnamefont
  {H.}~\bibnamefont {Häffner}}, \bibinfo {author} {\bibfnamefont
  {V.}~\bibnamefont {Kaushal}}, \bibinfo {author} {\bibfnamefont
  {T.}~\bibnamefont {Ruster}}, \bibinfo {author} {\bibfnamefont
  {C.}~\bibnamefont {Warschburger}}, \bibinfo {author} {\bibfnamefont
  {H.}~\bibnamefont {Kaufmann}}, \bibinfo {author} {\bibfnamefont {U.~G.}\
  \bibnamefont {Poschinger}}, \bibinfo {author} {\bibfnamefont
  {F.}~\bibnamefont {Schmidt-Kaler}}, \bibinfo {author} {\bibfnamefont
  {P.}~\bibnamefont {Schindler}}, \bibinfo {author} {\bibfnamefont
  {T.}~\bibnamefont {Monz}}, \ and\ \bibinfo {author} {\bibfnamefont
  {R.}~\bibnamefont {Blatt}},\ }\href@noop {} {\bibfield  {journal} {\bibinfo
  {journal} {Review of Scientific Instruments}\ }\textbf {\bibinfo {volume}
  {87}},\ \bibinfo {pages} {113103} (\bibinfo {year} {2016})}\BibitemShut
  {NoStop}%
\bibitem [{\citenamefont {Jackson}(1999)}]{Jackson_1999}%
  \BibitemOpen
  \bibfield  {author} {\bibinfo {author} {\bibfnamefont {J.~D.}\ \bibnamefont
  {Jackson}},\ }\href@noop {} {\emph {\bibinfo {title} {Classical
  electrodynamics}}}\ (\bibinfo  {publisher} {AAPT},\ \bibinfo {year}
  {1999})\BibitemShut {NoStop}%
\bibitem [{\citenamefont {Home}\ and\ \citenamefont
  {Steane}(2004)}]{Home_QuantumInfoComp_2004}%
  \BibitemOpen
  \bibfield  {author} {\bibinfo {author} {\bibfnamefont {J.}~\bibnamefont
  {Home}}\ and\ \bibinfo {author} {\bibfnamefont {A.}~\bibnamefont {Steane}},\
  }\href@noop {} {\bibfield  {journal} {\bibinfo  {journal} {Quantum
  information \& computation}\ }\textbf {\bibinfo {volume} {6}} (\bibinfo
  {year} {2004})}\BibitemShut {NoStop}%
\bibitem [{\citenamefont {Nizamani}\ and\ \citenamefont
  {Hensinger}(2012)}]{Nizamani_ApplPhysB_2012}%
  \BibitemOpen
  \bibfield  {author} {\bibinfo {author} {\bibfnamefont {A.~H.}\ \bibnamefont
  {Nizamani}}\ and\ \bibinfo {author} {\bibfnamefont {W.~K.}\ \bibnamefont
  {Hensinger}},\ }\href@noop {} {\bibfield  {journal} {\bibinfo  {journal}
  {Applied Physics B}\ }\textbf {\bibinfo {volume} {106}},\ \bibinfo {pages}
  {327} (\bibinfo {year} {2012})}\BibitemShut {NoStop}%
\bibitem [{\citenamefont {Eble}\ \emph {et~al.}(2010)\citenamefont {Eble},
  \citenamefont {Ulm}, \citenamefont {Zahariev}, \citenamefont
  {Schmidt-Kaler},\ and\ \citenamefont {Singer}}]{Eble_OptSocofAmerica_2010}%
  \BibitemOpen
  \bibfield  {author} {\bibinfo {author} {\bibfnamefont {J.}~\bibnamefont
  {Eble}}, \bibinfo {author} {\bibfnamefont {S.}~\bibnamefont {Ulm}}, \bibinfo
  {author} {\bibfnamefont {P.}~\bibnamefont {Zahariev}}, \bibinfo {author}
  {\bibfnamefont {F.}~\bibnamefont {Schmidt-Kaler}}, \ and\ \bibinfo {author}
  {\bibfnamefont {K.}~\bibnamefont {Singer}},\ }\href@noop {} {\bibfield
  {journal} {\bibinfo  {journal} {JOSA B}\ }\textbf {\bibinfo {volume} {27}},\
  \bibinfo {pages} {A99} (\bibinfo {year} {2010})}\BibitemShut {NoStop}%
\bibitem [{\citenamefont {Ruster}\ \emph {et~al.}(2014)\citenamefont {Ruster},
  \citenamefont {Warschburger}, \citenamefont {Kaufmann}, \citenamefont
  {Schmiegelow}, \citenamefont {Walther}, \citenamefont {Hettrich},
  \citenamefont {Pfister}, \citenamefont {Kaushal}, \citenamefont
  {Schmidt-Kaler},\ and\ \citenamefont {Poschinger}}]{Ruster_PhysRevA_2014}%
  \BibitemOpen
  \bibfield  {author} {\bibinfo {author} {\bibfnamefont {T.}~\bibnamefont
  {Ruster}}, \bibinfo {author} {\bibfnamefont {C.}~\bibnamefont
  {Warschburger}}, \bibinfo {author} {\bibfnamefont {H.}~\bibnamefont
  {Kaufmann}}, \bibinfo {author} {\bibfnamefont {C.~T.}\ \bibnamefont
  {Schmiegelow}}, \bibinfo {author} {\bibfnamefont {A.}~\bibnamefont
  {Walther}}, \bibinfo {author} {\bibfnamefont {M.}~\bibnamefont {Hettrich}},
  \bibinfo {author} {\bibfnamefont {A.}~\bibnamefont {Pfister}}, \bibinfo
  {author} {\bibfnamefont {V.}~\bibnamefont {Kaushal}}, \bibinfo {author}
  {\bibfnamefont {F.}~\bibnamefont {Schmidt-Kaler}}, \ and\ \bibinfo {author}
  {\bibfnamefont {U.~G.}\ \bibnamefont {Poschinger}},\ }\href {\doibase
  10.1103/PhysRevA.90.033410} {\bibfield  {journal} {\bibinfo  {journal} {Phys.
  Rev. A}\ }\textbf {\bibinfo {volume} {90}},\ \bibinfo {pages} {033410}
  (\bibinfo {year} {2014})}\BibitemShut {NoStop}%
\bibitem [{\citenamefont {Bruzewicz}\ \emph {et~al.}(2015)\citenamefont
  {Bruzewicz}, \citenamefont {Sage},\ and\ \citenamefont
  {Chiaverini}}]{Bruzewicz_PhysRevA_2015}%
  \BibitemOpen
  \bibfield  {author} {\bibinfo {author} {\bibfnamefont {C.~D.}\ \bibnamefont
  {Bruzewicz}}, \bibinfo {author} {\bibfnamefont {J.~M.}\ \bibnamefont {Sage}},
  \ and\ \bibinfo {author} {\bibfnamefont {J.}~\bibnamefont {Chiaverini}},\
  }\href {\doibase 10.1103/PhysRevA.91.041402} {\bibfield  {journal} {\bibinfo
  {journal} {Phys. Rev. A}\ }\textbf {\bibinfo {volume} {91}},\ \bibinfo
  {pages} {041402} (\bibinfo {year} {2015})}\BibitemShut {NoStop}%
\bibitem [{\citenamefont {Singer}\ \emph {et~al.}(2010)\citenamefont {Singer},
  \citenamefont {Poschinger}, \citenamefont {Murphy}, \citenamefont {Ivanov},
  \citenamefont {Ziesel}, \citenamefont {Calarco},\ and\ \citenamefont
  {Schmidt-Kaler}}]{Singer_RevModPhys_2009}%
  \BibitemOpen
  \bibfield  {author} {\bibinfo {author} {\bibfnamefont {K.}~\bibnamefont
  {Singer}}, \bibinfo {author} {\bibfnamefont {U.}~\bibnamefont {Poschinger}},
  \bibinfo {author} {\bibfnamefont {M.}~\bibnamefont {Murphy}}, \bibinfo
  {author} {\bibfnamefont {P.}~\bibnamefont {Ivanov}}, \bibinfo {author}
  {\bibfnamefont {F.}~\bibnamefont {Ziesel}}, \bibinfo {author} {\bibfnamefont
  {T.}~\bibnamefont {Calarco}}, \ and\ \bibinfo {author} {\bibfnamefont
  {F.}~\bibnamefont {Schmidt-Kaler}},\ }\href@noop {} {\bibfield  {journal}
  {\bibinfo  {journal} {Reviews of Modern Physics}\ }\textbf {\bibinfo {volume}
  {82}},\ \bibinfo {pages} {2609} (\bibinfo {year} {2010})}\BibitemShut
  {NoStop}%
\bibitem [{\citenamefont {Leibfried}\ \emph {et~al.}(2003)\citenamefont
  {Leibfried}, \citenamefont {Blatt}, \citenamefont {Monroe},\ and\
  \citenamefont {Wineland}}]{Leibfried_RevModPhys_2003}%
  \BibitemOpen
  \bibfield  {author} {\bibinfo {author} {\bibfnamefont {D.}~\bibnamefont
  {Leibfried}}, \bibinfo {author} {\bibfnamefont {R.}~\bibnamefont {Blatt}},
  \bibinfo {author} {\bibfnamefont {C.}~\bibnamefont {Monroe}}, \ and\ \bibinfo
  {author} {\bibfnamefont {D.}~\bibnamefont {Wineland}},\ }\href@noop {}
  {\bibfield  {journal} {\bibinfo  {journal} {Reviews of Modern Physics}\
  }\textbf {\bibinfo {volume} {75}},\ \bibinfo {pages} {281} (\bibinfo {year}
  {2003})}\BibitemShut {NoStop}%
\bibitem [{\citenamefont {Keller}\ \emph {et~al.}(2015)\citenamefont {Keller},
  \citenamefont {Partner}, \citenamefont {Burgermeister},\ and\ \citenamefont
  {Mehlst{\"a}ubler}}]{Keller_JAP_2015}%
  \BibitemOpen
  \bibfield  {author} {\bibinfo {author} {\bibfnamefont {J.}~\bibnamefont
  {Keller}}, \bibinfo {author} {\bibfnamefont {H.}~\bibnamefont {Partner}},
  \bibinfo {author} {\bibfnamefont {T.}~\bibnamefont {Burgermeister}}, \ and\
  \bibinfo {author} {\bibfnamefont {T.}~\bibnamefont {Mehlst{\"a}ubler}},\
  }\href@noop {} {\bibfield  {journal} {\bibinfo  {journal} {J. of Appl.
  Phys.}\ }\textbf {\bibinfo {volume} {118}},\ \bibinfo {pages} {104501}
  (\bibinfo {year} {2015})}\BibitemShut {NoStop}%
\bibitem [{\citenamefont {Berkeland}\ \emph {et~al.}(1998)\citenamefont
  {Berkeland}, \citenamefont {Miller}, \citenamefont {Bergquist}, \citenamefont
  {Itano},\ and\ \citenamefont {Wineland}}]{Berkeland_JApplPhys_1998}%
  \BibitemOpen
  \bibfield  {author} {\bibinfo {author} {\bibfnamefont {D.~J.}\ \bibnamefont
  {Berkeland}}, \bibinfo {author} {\bibfnamefont {J.~D.}\ \bibnamefont
  {Miller}}, \bibinfo {author} {\bibfnamefont {J.~C.}\ \bibnamefont
  {Bergquist}}, \bibinfo {author} {\bibfnamefont {W.~M.}\ \bibnamefont
  {Itano}}, \ and\ \bibinfo {author} {\bibfnamefont {D.~J.}\ \bibnamefont
  {Wineland}},\ }\href {\doibase 10.1063/1.367318} {\bibfield  {journal}
  {\bibinfo  {journal} {J. of Appl. Phys.}\ }\textbf {\bibinfo {volume} {83}},\
  \bibinfo {pages} {5025} (\bibinfo {year} {1998})}\BibitemShut {NoStop}%
\bibitem [{\citenamefont {Home}\ \emph {et~al.}(2011)\citenamefont {Home},
  \citenamefont {Hanneke}, \citenamefont {Jost}, \citenamefont {Leibfried},\
  and\ \citenamefont {Wineland}}]{Home_NJP_2011}%
  \BibitemOpen
  \bibfield  {author} {\bibinfo {author} {\bibfnamefont {J.}~\bibnamefont
  {Home}}, \bibinfo {author} {\bibfnamefont {D.}~\bibnamefont {Hanneke}},
  \bibinfo {author} {\bibfnamefont {J.}~\bibnamefont {Jost}}, \bibinfo {author}
  {\bibfnamefont {D.}~\bibnamefont {Leibfried}}, \ and\ \bibinfo {author}
  {\bibfnamefont {D.}~\bibnamefont {Wineland}},\ }\href@noop {} {\bibfield
  {journal} {\bibinfo  {journal} {New J. of Phys.}\ }\textbf {\bibinfo {volume}
  {13}},\ \bibinfo {pages} {073026} (\bibinfo {year} {2011})}\BibitemShut
  {NoStop}%
\bibitem [{\citenamefont {Chou}\ \emph {et~al.}(2010)\citenamefont {Chou},
  \citenamefont {Hume}, \citenamefont {Koelemeij}, \citenamefont {Wineland},\
  and\ \citenamefont {Rosenband}}]{Chou_PRL_2010}%
  \BibitemOpen
  \bibfield  {author} {\bibinfo {author} {\bibfnamefont {C.~W.}\ \bibnamefont
  {Chou}}, \bibinfo {author} {\bibfnamefont {D.~B.}\ \bibnamefont {Hume}},
  \bibinfo {author} {\bibfnamefont {J.~C.~J.}\ \bibnamefont {Koelemeij}},
  \bibinfo {author} {\bibfnamefont {D.~J.}\ \bibnamefont {Wineland}}, \ and\
  \bibinfo {author} {\bibfnamefont {T.}~\bibnamefont {Rosenband}},\ }\href
  {\doibase 10.1103/PhysRevLett.104.070802} {\bibfield  {journal} {\bibinfo
  {journal} {Phys. Rev. Lett.}\ }\textbf {\bibinfo {volume} {104}},\ \bibinfo
  {pages} {070802} (\bibinfo {year} {2010})}\BibitemShut {NoStop}%
\bibitem [{\citenamefont {Brownnutt}\ \emph {et~al.}(2015)\citenamefont
  {Brownnutt}, \citenamefont {Kumph}, \citenamefont {Rabl},\ and\ \citenamefont
  {Blatt}}]{Brownutt_RevModPhys_2015}%
  \BibitemOpen
  \bibfield  {author} {\bibinfo {author} {\bibfnamefont {M.}~\bibnamefont
  {Brownnutt}}, \bibinfo {author} {\bibfnamefont {M.}~\bibnamefont {Kumph}},
  \bibinfo {author} {\bibfnamefont {P.}~\bibnamefont {Rabl}}, \ and\ \bibinfo
  {author} {\bibfnamefont {R.}~\bibnamefont {Blatt}},\ }\href {\doibase
  10.1103/RevModPhys.87.1419} {\bibfield  {journal} {\bibinfo  {journal} {Rev.
  Mod. Phys.}\ }\textbf {\bibinfo {volume} {87}},\ \bibinfo {pages} {1419}
  (\bibinfo {year} {2015})}\BibitemShut {NoStop}%
\bibitem [{\citenamefont {Murphy}(2012)}]{Murphy_2012}%
  \BibitemOpen
  \bibfield  {author} {\bibinfo {author} {\bibfnamefont {K.~P.}\ \bibnamefont
  {Murphy}},\ }\href@noop {} {\emph {\bibinfo {title} {Machine learning: a
  probabilistic perspective}}}\ (\bibinfo  {publisher} {MIT press},\ \bibinfo
  {address} {Cambridge, MA},\ \bibinfo {year} {2012})\BibitemShut {NoStop}%
\bibitem [{\citenamefont {Diedrich}\ \emph {et~al.}(1989)\citenamefont
  {Diedrich}, \citenamefont {Bergquist}, \citenamefont {Itano},\ and\
  \citenamefont {Wineland}}]{Diedrich_RhysRevLett_1989}%
  \BibitemOpen
  \bibfield  {author} {\bibinfo {author} {\bibfnamefont {F.}~\bibnamefont
  {Diedrich}}, \bibinfo {author} {\bibfnamefont {J.~C.}\ \bibnamefont
  {Bergquist}}, \bibinfo {author} {\bibfnamefont {W.~M.}\ \bibnamefont
  {Itano}}, \ and\ \bibinfo {author} {\bibfnamefont {D.~J.}\ \bibnamefont
  {Wineland}},\ }\href {\doibase 10.1103/PhysRevLett.62.403} {\bibfield
  {journal} {\bibinfo  {journal} {Phys. Rev. Lett.}\ }\textbf {\bibinfo
  {volume} {62}},\ \bibinfo {pages} {403} (\bibinfo {year} {1989})}\BibitemShut
  {NoStop}%
\bibitem [{\citenamefont {Nielsen}\ and\ \citenamefont
  {Chuang}(2011)}]{Nielsen_2011}%
  \BibitemOpen
  \bibfield  {author} {\bibinfo {author} {\bibfnamefont {M.~A.}\ \bibnamefont
  {Nielsen}}\ and\ \bibinfo {author} {\bibfnamefont {I.~L.}\ \bibnamefont
  {Chuang}},\ }\href@noop {} {\emph {\bibinfo {title} {Quantum Computation and
  Quantum Information: 10th Anniversary Edition}}},\ \bibinfo {edition} {10th}\
  ed.\ (\bibinfo  {publisher} {Cambridge University Press},\ \bibinfo {address}
  {New York, NY, USA},\ \bibinfo {year} {2011})\BibitemShut {NoStop}%
\bibitem [{\citenamefont {Marquet}\ \emph {et~al.}(2003)\citenamefont
  {Marquet}, \citenamefont {Schmidt-Kaler},\ and\ \citenamefont
  {James}}]{Marquet_ApplPhysB_2003}%
  \BibitemOpen
  \bibfield  {author} {\bibinfo {author} {\bibfnamefont {C.}~\bibnamefont
  {Marquet}}, \bibinfo {author} {\bibfnamefont {F.}~\bibnamefont
  {Schmidt-Kaler}}, \ and\ \bibinfo {author} {\bibfnamefont {D.~F.}\
  \bibnamefont {James}},\ }\href@noop {} {\bibfield  {journal} {\bibinfo
  {journal} {Applied Physics B}\ }\textbf {\bibinfo {volume} {76}},\ \bibinfo
  {pages} {199} (\bibinfo {year} {2003})}\BibitemShut {NoStop}%
\bibitem [{\citenamefont {Tan}\ \emph {et~al.}(2015)\citenamefont {Tan},
  \citenamefont {Gaebler}, \citenamefont {Lin}, \citenamefont {Wan},
  \citenamefont {Bowler}, \citenamefont {Leibfried},\ and\ \citenamefont
  {Wineland}}]{Tan_Nature_2015}%
  \BibitemOpen
  \bibfield  {author} {\bibinfo {author} {\bibfnamefont {T.~R.}\ \bibnamefont
  {Tan}}, \bibinfo {author} {\bibfnamefont {J.~P.}\ \bibnamefont {Gaebler}},
  \bibinfo {author} {\bibfnamefont {Y.}~\bibnamefont {Lin}}, \bibinfo {author}
  {\bibfnamefont {Y.}~\bibnamefont {Wan}}, \bibinfo {author} {\bibfnamefont
  {R.}~\bibnamefont {Bowler}}, \bibinfo {author} {\bibfnamefont
  {D.}~\bibnamefont {Leibfried}}, \ and\ \bibinfo {author} {\bibfnamefont
  {D.~J.}\ \bibnamefont {Wineland}},\ }\href@noop {} {\bibfield  {journal}
  {\bibinfo  {journal} {Nature}\ }\textbf {\bibinfo {volume} {528}},\ \bibinfo
  {pages} {380} (\bibinfo {year} {2015})}\BibitemShut {NoStop}%
\bibitem [{\citenamefont {Hume}\ \emph {et~al.}(2009)\citenamefont {Hume},
  \citenamefont {Chou}, \citenamefont {Rosenband},\ and\ \citenamefont
  {Wineland}}]{Hume_PhysRevA_2009}%
  \BibitemOpen
  \bibfield  {author} {\bibinfo {author} {\bibfnamefont {D.}~\bibnamefont
  {Hume}}, \bibinfo {author} {\bibfnamefont {C.-W.}\ \bibnamefont {Chou}},
  \bibinfo {author} {\bibfnamefont {T.}~\bibnamefont {Rosenband}}, \ and\
  \bibinfo {author} {\bibfnamefont {D.~J.}\ \bibnamefont {Wineland}},\
  }\href@noop {} {\bibfield  {journal} {\bibinfo  {journal} {Physical Review
  A}\ }\textbf {\bibinfo {volume} {80}},\ \bibinfo {pages} {052302} (\bibinfo
  {year} {2009})}\BibitemShut {NoStop}%
\bibitem [{\citenamefont {Erhard}\ \emph {et~al.}(2019)\citenamefont {Erhard},
  \citenamefont {Wallman}, \citenamefont {Postler}, \citenamefont {Meth},
  \citenamefont {Stricker}, \citenamefont {Martinez}, \citenamefont
  {Schindler}, \citenamefont {Monz}, \citenamefont {Emerson},\ and\
  \citenamefont {Blatt}}]{Erhard_arXiv_2019}%
  \BibitemOpen
  \bibfield  {author} {\bibinfo {author} {\bibfnamefont {A.}~\bibnamefont
  {Erhard}}, \bibinfo {author} {\bibfnamefont {J.~J.}\ \bibnamefont {Wallman}},
  \bibinfo {author} {\bibfnamefont {L.}~\bibnamefont {Postler}}, \bibinfo
  {author} {\bibfnamefont {M.}~\bibnamefont {Meth}}, \bibinfo {author}
  {\bibfnamefont {R.}~\bibnamefont {Stricker}}, \bibinfo {author}
  {\bibfnamefont {E.~A.}\ \bibnamefont {Martinez}}, \bibinfo {author}
  {\bibfnamefont {P.}~\bibnamefont {Schindler}}, \bibinfo {author}
  {\bibfnamefont {T.}~\bibnamefont {Monz}}, \bibinfo {author} {\bibfnamefont
  {J.}~\bibnamefont {Emerson}}, \ and\ \bibinfo {author} {\bibfnamefont
  {R.}~\bibnamefont {Blatt}},\ }\href@noop {} {\bibfield  {journal} {\bibinfo
  {journal} {arXiv preprint arXiv:1902.08543}\ } (\bibinfo {year}
  {2019})}\BibitemShut {NoStop}%
\bibitem [{\citenamefont {Finkel}\ and\ \citenamefont
  {Kelley}(2004)}]{Finkel_OptOnl_2004}%
  \BibitemOpen
  \bibfield  {author} {\bibinfo {author} {\bibfnamefont {D.~E.}\ \bibnamefont
  {Finkel}}\ and\ \bibinfo {author} {\bibfnamefont {C.}~\bibnamefont
  {Kelley}},\ }\href@noop {} {\bibfield  {journal} {\bibinfo  {journal}
  {Optimization Online}\ }\textbf {\bibinfo {volume} {14}},\ \bibinfo {pages}
  {1} (\bibinfo {year} {2004})}\BibitemShut {NoStop}%
\bibitem [{\citenamefont {Morigi}\ \emph {et~al.}(2000)\citenamefont {Morigi},
  \citenamefont {Eschner},\ and\ \citenamefont
  {Keitel}}]{Morigi_PhysRevLett_2000}%
  \BibitemOpen
  \bibfield  {author} {\bibinfo {author} {\bibfnamefont {G.}~\bibnamefont
  {Morigi}}, \bibinfo {author} {\bibfnamefont {J.}~\bibnamefont {Eschner}}, \
  and\ \bibinfo {author} {\bibfnamefont {C.~H.}\ \bibnamefont {Keitel}},\
  }\href {\doibase 10.1103/PhysRevLett.85.4458} {\bibfield  {journal} {\bibinfo
   {journal} {Phys. Rev. Lett.}\ }\textbf {\bibinfo {volume} {85}},\ \bibinfo
  {pages} {4458} (\bibinfo {year} {2000})}\BibitemShut {NoStop}%
\bibitem [{\citenamefont {Ejtemaee}\ and\ \citenamefont
  {Haljan}(2017)}]{Ejtemaee_PhysRevLett_2017}%
  \BibitemOpen
  \bibfield  {author} {\bibinfo {author} {\bibfnamefont {S.}~\bibnamefont
  {Ejtemaee}}\ and\ \bibinfo {author} {\bibfnamefont {P.~C.}\ \bibnamefont
  {Haljan}},\ }\href {\doibase 10.1103/PhysRevLett.119.043001} {\bibfield
  {journal} {\bibinfo  {journal} {Phys. Rev. Lett.}\ }\textbf {\bibinfo
  {volume} {119}},\ \bibinfo {pages} {043001} (\bibinfo {year}
  {2017})}\BibitemShut {NoStop}%
\bibitem [{\citenamefont {Bunse-Gerstner}(1984)}]{Bunse_gerstner_1984}%
  \BibitemOpen
  \bibfield  {author} {\bibinfo {author} {\bibfnamefont {A.}~\bibnamefont
  {Bunse-Gerstner}},\ }\href {\doibase
  https://doi.org/10.1016/0024-3795(84)90203-9} {\bibfield  {journal} {\bibinfo
   {journal} {Linear Algebra and its Applications}\ }\textbf {\bibinfo {volume}
  {58}},\ \bibinfo {pages} {43} (\bibinfo {year} {1984})}\BibitemShut {NoStop}%
\end{thebibliography}
